\documentclass[letterpaper,twocolumn,10pt]{article}
\usepackage{usenix-2020-09,epsfig,endnotes}
\usepackage{subfigure}
\usepackage{amsmath}
\usepackage{xspace}
\usepackage{circledsteps}
\usepackage{authblk}

\def\Snospace~{\S{}}

\usepackage[linesnumbered,ruled,vlined]{algorithm2e}
\newcommand{\sys}{Swarm\xspace}

\newcommand{\sch}{ByteScheduler\xspace}

\begin{document}

%\title{BytePCDN: Bridging the Gap Between CDN Efficiency and P2P Flexibility/ From CDN to PCDN: A Journey of Decentralized Video Streaming Optimization}
\title{Swarm: Cost-Efficient Video Content Distribution with a Peer-to-Peer System}
% \author[1 4]{Dehui Wei}
% \author[1 2 *]{Jiao Zhang}
% \author[3]{Haozhe Li}
% \author[3]{Zhichen Xue}
% \author[3]{Yajie Peng}
% \author[3]{\\Xiaofei Pang}
% \author[3]{Rui Han}
% \author[3]{Yan Ma}
% \author[4]{Jialin Li}
% \affil[1]{State Key Laboratory of Networking and
% 	Switching Technology, BUPT}
%  \affil[2]{Purple Mountain Laboratories}
%  \affil[3]{ByteDance, Beijing}
%  \affil[4]{School of Computing, National University of Singapore}
%\renewcommand{\thefootnote}{\fnsymbol{footnote}}
\author{
{ Dehui Wei$^{1,4}$,  Jiao Zhang $^{1,2,*}$,  Haozhe Li$^3$, Zhichen Xue$^3$, Yajie Peng$^3$,  

Xiaofei Pang$^3$, Rui Han$^3$, Yan Ma$^3$, and Jialin Li$^4$}

$^1$\textit{Beijing University of Posts and Communications, }$^2$\textit{Purple Mountain Laboratories, }

$^3$\textit{ByteDance, Beijing, $^4$School of Computing, }\textit{National University of Singapore}
} % end author
 
% \author[1 4]{Dehui Wei}
% \author[1 2 *]{Jiao Zhang}
% \author[3]{Haozhe Li}
% \author[3]{Zhichen Xue}
% \author[3]{Yajie Peng}
% \author[3]{\\Xiaofei Pang}
% \author[3]{Rui Han}
% \author[3]{Yan Ma}
% \author[4]{Jialin Li}
% \affil[1]{State Key Laboratory of Networking and
% 	Switching Technology, BUPT}
%  \affil[2]{Purple Mountain Laboratories}
%  \affil[3]{ByteDance, Beijing}
%  \affil[4]{School of Computing, National University of Singapore}
\maketitle
\pagestyle{plain}

\begin{abstract}

As ByteDance's business expands, the substantial infrastructure expenses associated with centralized Content Delivery Network (CDN) networks have rendered content distribution costs prohibitively high. In response, we embarked on exploring a peer-to-peer (P2P) network as a promising solution to alleviate the escalating costs of content distribution. However, the decentralized nature of P2P often introduces performance challenges, given the diversity and dispersion of peer devices.
This study introduces \textit{\sys}, ByteDance's innovative hybrid system for video streaming. \sys seamlessly integrates the robustness of a conventional CDN with the cost-efficiency of a decentralized P2P network. Its primary aim is to provide users with reliable streaming quality while minimizing traffic expenses. To achieve this, \sys employs a centralized control plane comprised of a tracker cluster, overseeing a data plane with numerous edge residual resources. The tracker also takes on the responsibility of mapping clients to servers.
Addressing the performance disparities among individual peer servers, \sys utilizes our proprietary multipath parallel transmission method for communication between clients and peer servers. Operating stably for six years, \sys now manages over a hundred thousand peer servers, serving nearly a hundred million users daily and saving the company hundreds of millions of RMB annually. Experimental results affirm that, while significantly cutting costs, \sys performs on par with traditional CDNs.

\end{abstract}

\section{Introduction}
At ByteDance, our content delivery network (CDN)\cite{peng2004cdn} plays a critical role in serving billions of global users across our wide range of products and services.
As our business continues to grow, we are forced to increase the capacity of our CDN infrastructure to meet the surging demand. 
This continuous expansion comes with a steep hardware, network, and management cost in our centralized CDN data centers.
Centralization also leads to poor agility and flexibility\cite{yin2009design,plagemann2006content}.
Even though our user traffic exhibits high temporal variance, we have to over provision our data centers for peak traffic.
These compounding factors have led to an unsustainable increase in the average device and bandwidth 
expenses of our CDN infrastructure.

%Therefore, it's imperative for us to devise strategies to trim on-demand video streaming costs without compromising the user experience, ensuring we persist in delivering premium services.

A decentralized peer-to-peer (P2P) network\cite{farahani2022hybrid,jiang2009efficient} offers a promising solution to reduce the cost of content distribution.
%Within a P2P framework, each node operates on an egalitarian principle: a node can function both as a downloader, procuring video data from other nodes, and as an uploader, supplying video data to peers.
In contrast to centralized deployments, a P2P network harnesses \emph{underutilized resources} scattered across global devices, including set-top boxes, smart home gadgets, dormant servers, and more.
These devices incur significantly lower per-unit hardware, bandwidth, electricity, and management cost than their data center counterparts\cite{zhang2014unreeling,kang2010hybrid}. 
More importantly, P2P enables \emph{resource elasticity}\cite{zhang2018proactive}.
The network can scale up or down dynamically based on demand, and only pays for the resources required to serve the actual traffic.
%By harnessing these underutilized resources in a P2P manner, we can achieve a lower unit cost, with traffic costs rising linearly as demand increases.

Decentralization, however, comes at a performance cost\cite{ma2021locality}.
Peer devices in a P2P network are highly heterogeneous: They exhibit large variance in their network bandwidth, storage capacity, and computing capabilities;
the network experiences frequent churns, where devices join the leave the system at high rate;
peers are dispersed across large geographic locations, regional networks, and ISPs, without any central management;
most P2P devices are behind a private network, requiring lengthy Network Address Translation (NAT)\cite{egevang1994rfc1631,egevang1994ip} penetration during connection setup.
As a consequence, decentralization presents formidable challenges to provide the strict, real-time content delivery requirements of our services\cite{farahani2022richter}.

%In contrast to the hierarchical structure of CDNs that use dedicated servers, these cost-effective devices, termed \textit{`peer servers'}, operate on a flat architectural framework. 
%These peer servers are numerous and dispersed across the nation, which makes centralized management tricky due to their varied and widespread locations. 
%Moreover, a majority of these machines operate within LAN (Local Area Network), necessitating Network Address Translation (NAT) penetration, which can lead to prolonged start-up times for content playback.

%Some data: Currently, BytePCDN boasts over 100,000 server nodes, spanning across all provinces in China.

In this paper, we present \sys, ByteDance's hybrid content distribution system for video streaming.
To deliver reliable video streaming quality to our users while reducing infrastructure cost, \sys combines a traditional CDN with a decentralized, P2P distribution network (PCDN).
The CDN serves three main purposes:
it provides a video content ``backend storage'' for the PCDN;
its more predictable performance is leveraged to reduce video loading time;
it serves as a fallback option when the quality of service of PCDN drops below Service Level Objective (SLOs).
\sys, however, does not provision the CDN to serve peak user traffic.
Instead, it relies on the PCDN for cost-effective and elastic capacity provisioning.

The main technical contribution of our paper is the design of the PCDN system.
Our PCDN employs a pool of underutilized low-cost devices outside our managed data centers and networks.
The system delivers video streaming quality \emph{on par} with dedicated CDN infrastructure, despite all the shortcomings of a P2P network discussed above.
To achieve this challenging goal, we carefully divide the content distribution responsibility between a fast but simple data transmission plane, and a centrally managed decision making engine.
This centralized control plane consists of a cluster of \emph{trackers} running in our data centers.
Centralization enables the trackers to have a global view of P2P peer server status, content location, network traffic, user demand, and service requirements.
Trackers can therefore make fine-grained and highly optimized resource allocation, content distribution, and client-to-server mapping decisions.

The vast amount of P2P peer servers power the PCDN data plane.
As individual peer servers may not have the computational power or network bandwidth to deliver the required video streaming quality, we design a novel Multi-site Parallel Downloading (MPD) scheme.
Specifically, client devices establish connections to multiple peer servers caching the requested video, and schedule concurrent video transmission from all connected peers.
Our MPD transmission scheme aggregates computation and network capacity from a pool of peer servers, ensuring video downloading performance that meets our requirements.

%In this paper, we present a hybrid content distribution system that fuses centralized and decentralized paradigms — christened the PCDN system, a stalwart that has reliably powered ByteDance for over six years.
%PCDN's chief objective is to employ budget-friendly devices and networks to consistently deliver top-notch video streaming quality to our users.
%One of the challenges faced by this decentralized networks with inexpensive computing resources is the delivery of content with the same quality and reliability as dedicated CDN servers.
%One solution to this challenge is the Multi-site Parallel Downloading (MPD) approach, which bridges this performance gap. However, NAT penetration introduces latency during playback initiation.
%To circumvent this, PCDN initiates streaming from the CDN. Once the PCDN connection is established and ready, the download switches over to PCDN.
%Distributing video data efficiently is crucial; thus, PCDN adopts a centralized scheduling mechanism based on the popularity of videos.
%This centralized control plane is known as the Tracker.
%The Tracker offers a holistic view of the distribution network, understanding where specific video data is located and monitoring the status of each peer server.
\begin{figure}
    \centering
    \includegraphics[width=0.38\textwidth]{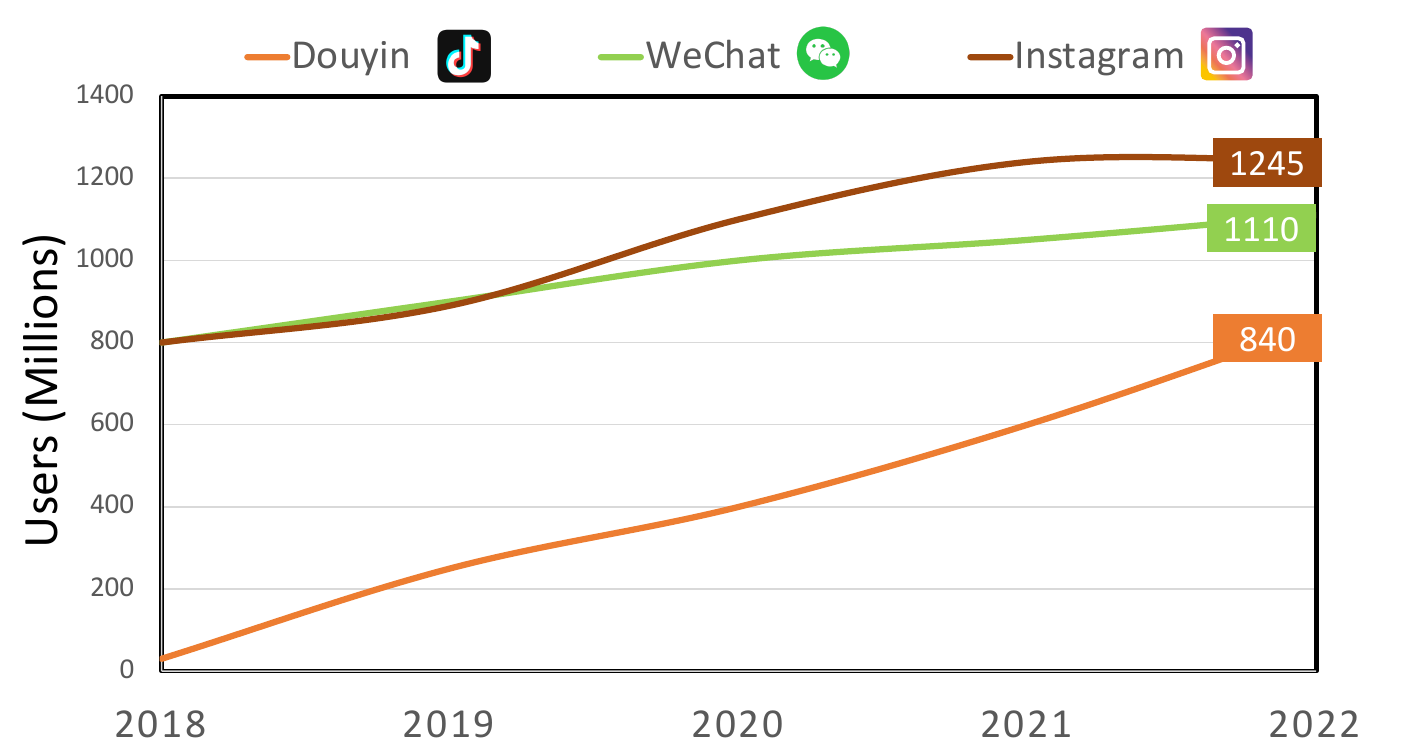}
    \caption{Growth in user numbers for various video applications in recent years.}
    \label{user}
\end{figure}
\sys has served as ByteDance's main content distribution engine for over six years.
It has grown to 100,000 peer servers distributed across all provinces in mainland China.
Serving close to a hundred million daily users, \sys processes more than ten billion daily playbacks and handles tens of terabytes of video traffic per day.
Our strategy of embracing decentralization has led to an annual cost saving of hundreds of millions of RMB for the company.
More importantly, cost-efficiency of out PCDN design did not compromise user experience.
During the 2022 World Cup, \sys successfully managed the exponential surge in user demand, delivering uninterrupted live broadcasts of the games without any major incidents.

\begin{figure}
    \centering
    \includegraphics[width=0.3\textwidth]{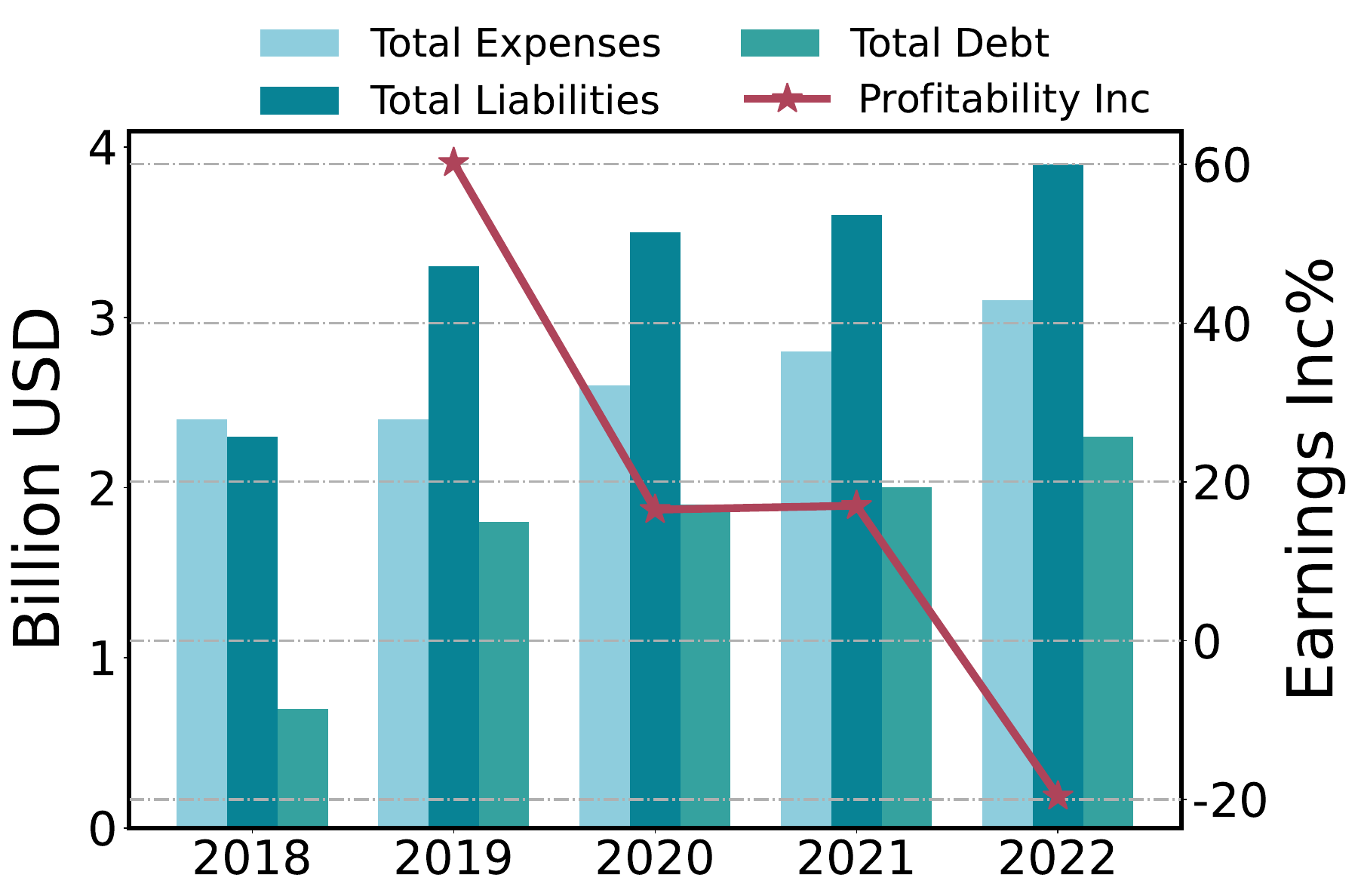}
    \caption{Akamai investment expenditure and profit growth rate.}
    \label{akamai}
\end{figure}

\section{Background and Motivation}

Videos have emerged as the dominant content in today's internet landscape, as evidenced by \autoref{user} which shows the growth of various video applications in user count over the past few years.
Platforms such as Douyin and TikTok, in particular, have demonstrated an expedited growth pattern.
The growth in video popularity has resulted in rapid increase in video streaming network traffic.
This massive traffic surge has brought unprecedented challenges to the content delivery network infrastructure to continue delivering high-quality video streaming service.
%both CDN and Peer-to-peer networks are confronting distinct challenges.

\subsection{Content Distribution Networks}
\noindent\textbf{CDN Overview: } 
A CDN deploys dedicated server clusters across vast geographical locations.
CDNs scale content delivery capacity and enable users to fetch data from servers in close proximity\cite{adhikari2012unreeling}.
To maintain fast and reliable video transmission, CDN clusters deploy powerful rack servers and use dedicated commercial networks from large ISPs.
%Managed predominantly by ISPs, CDN clusters commonly use dedicated commercial networks, vital for maintaining high-quality transmission.
Each CDN integrates its unique DNS system to direct user requests for content\cite{otto2012content}.
The DNS service connects users to the nearest CDN server, optimizing performance.
%CDNs prioritize speed and reliability, prompting investment in top-tier servers.
While all CDNs use a hierarchical model to disseminate content to edge servers, their specific system architectures may differ\cite{xu2010toward}. Some, like Akamai\cite{de2010experimental}, prefer a fully distributed approach where edge servers are deployed globally. 
In contrast, companies such as Limelight\cite{canali2018designing} opt for a more centralized architecture, in which they build fewer but larger data centers that interface with ISPs. 
The decision on the server distribution largely hinges on a balance between bandwidth capacity and the associated costs.

\noindent\textbf{Challenge: }
%The escalating consumption of videos, along with the ensuing demand for a higher server count, has surged network traffic. 
This video demand boom has lead to substantial CDN traffic cost increase for content providers\cite{wang2023trafada}. 
Indicative of its soaring demand, the CDN market size is expected to soar to an estimated USD 1055.5 billion by 2032, with a remarkable CAGR of 18.31\% from 2023 to 2032\cite{hasanov2023content}.
A stark reflection of this trend is Douyin, which reports yearly traffic expenses surpassing 6 billion RMB. 
When CDNs hit throughput constraints, there's a pressing need to integrate more server points, which in turn amplifies construction costs.
Companies following a distributed approach, like Akamai, combat these challenges by augmenting the number of edge servers. 
This strategy inevitably leads to elevated operating costs, encompassing aspects like electricity, widespread content replication, and associated storage and bandwidth costs for traffic redirection.
\autoref{akamai} shows Akamai's spiraling annual expenses\cite{website:statistics}. 
Its dwindling profit margins have lead to recent announcement of a price increase.
On the flip side, while centralized CDNs benefit from reduced management costs, they grapple with an accelerated rate of unit price hikes in tandem with throughput escalation.
This is predominantly because each tier demands additional servers as traffic burgeons.

\subsection{Peer-to-Peer Network}
\noindent\textbf{P2P Overview}:
Peer-to-Peer (P2P) networks are recognized for their decentralized architecture and have gained popularity due to their distinctive mode of operation. In a P2P system, peers collaboratively create a distributed framework with the core objective of sharing content\cite{ansari2021score}. Each peer operates on an egalitarian basis, serving dually as both a consumer and a provider of resources. Interactions within a P2P system occur in two primary ways: users initiate search queries to locate desired content, and upon discovering the relevant objects, they proceed to download the content. Contrary to web and CDN systems, P2P systems primarily engage users in a non-interactive, batch-style content download\cite{saroiu2002analysis}.  A significant advantage of P2P networks is their cost-effectiveness; as demand scales, they do not necessitate the large infrastructure investments seen in CDNs.

\noindent\textbf{Challenge:}
While P2P networks offer notable advantages, they simultaneously introduce specific performance and management challenges. The majority of content-serving hosts in P2P systems, operated by end-users, often grapple with low availability and have comparatively slower network connections \cite{yang2002efficient}. Companies are limited in their capacity to directly intervene in users' storage and transmission methods within these networks and only can rely on algorithms for guidance. As specific content gains traction and popularity within the P2P system, a small number of peers can consume an enormous amount of bandwidth~\cite{saroiu2002analysis}. 
The inherent flexibility to swiftly introduce new nodes to counter this demand is absent, potentially leading to service disruptions. Further complicating matters is the dynamic nature of P2P networks, where nodes can join or depart at their discretion, introducing elements of unpredictability and potential bandwidth inconsistencies\cite{chun2005impact}. As such, an exclusive reliance on a decentralized P2P framework for video distribution raises concerns about ensuring a consistent quality of service. 

In summary: While centralized CDN systems come at a high cost, they ensure satisfactory content service quality.
In contrast, decentralized P2P systems, despite being cost-effective, prove to be inefficient and challenging to manage.

\subsection{Opportunity: Leverage P2P For Cost Reduction}
Can we exploit a P2P network to reduce our content delivery cost?
Our investigations have unveiled a treasure trove of underutilized resources nestled within household devices such as set-top boxes, smart home appliances, and occasionally idle servers. 
These devices, often geographically closer to users, represent an untapped potential for content storage and distribution. 
They also offer significant cost-saving opportunities compared to a centralized CDN infrastructure.
This cost-saving comes from two main factors:
\textit{1) Utilization of inexpensive resources}: Compared to data center rack servers, these devices have much lower hardware and utility cost.
Moreover, since these devices are not owned by ByteDance, we can tap into the under-utilized portion of their resources at a fraction of the overall cost.
\textit{2) Avoiding expensive dedicated networks}: Traditional CDNs spend significant portion of their infrastructure budget on building and maintaining dedicated networks.
These peer devices, however, are connected to public networks from commercial ISPs. 
The cost of their under-utilized network bandwidth (particularly for upload) is also a fraction of a large data center network.

\begin{figure}
    \centering
    \includegraphics[width=0.45\textwidth]{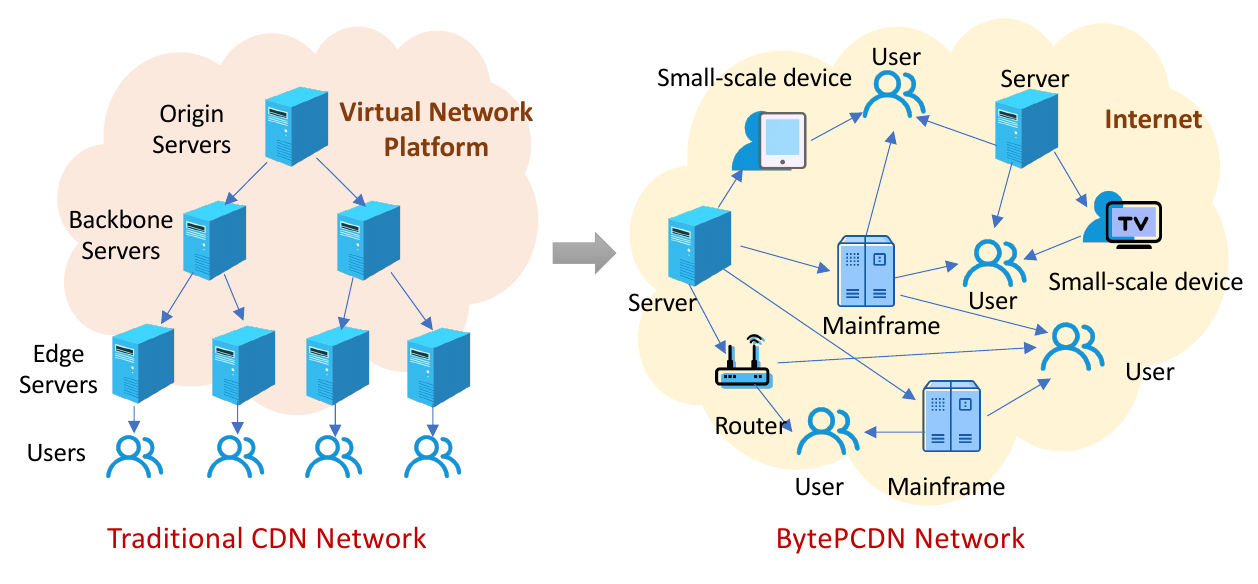}
    \caption{Differences between CDN and \sys architectures.}
    \label{CDN and PCDN.}
\end{figure}

Our P2P CDN (PCDN) infrastructure capitalizes on these latent resources to drive down the traffic-related costs associated with conventional CDNs. \autoref{CDN and PCDN.} illustrates the differences between the architectures of CDN and PCDN. 
As we introduce later in the paper, our PCDN addresses the inherent instability of these fault-prone, less powerful devicesthrough a centralized control platform to maintain video delivery quality. 
Consequently, PCDN manages to reduce the expenses associated with on-demand video streaming, all the while delivering user experience comparable to a traditional CDN.

Moreover, embodying a P2P structure, PCDN promotes a dynamic ecosystem where any peer server can seamlessly integrate into or exit from the system.
This fluidity ensures enhanced dynamic scaling, allowing traffic costs to align and grow proportionally to traffic demand.

%An in-depth analysis of user bandwidth and storage resources reveals that there is an abundance of underutilized assets. 
%We conduct an experiment to graphically represent the usage percentage and duration of these resources, thereby providing a quantitative view of the situation.

% Experiment: Graph depicting the usage percentage and duration of user bandwidth, as well as storage resource utilization.

% (Include a graph showing the distribution of resources among different devices and the amount of remaining resources available for utilization.)

\subsection{Design Challenges}
While using a P2P network has the potential to reduce content delivery cost, this cost-saving cannot come at the expense of user experience.
Simultaneously satisfying the affordability and performance requirements introduce the following design challenges. 
%Our commitment remains unwavering: to consistently deliver top-notch video quality and ensure an impeccable user experience, all the while ensuring sustainability and affordability. 
%This is our promise, as we strive to maintain high-caliber services for our global user base.

\textbf{Performance Limitations of Peer Servers}: In stark contrast to traditional CDNs, which benefit from dedicated servers, PCDN leverages inexpensive and idle resources from household devices. While cost-effective, these Peer Servers come with their own set of challenges. They tend to have just enough storage capacity, frequently face network bandwidth that's not only variable but also often inadequate, and they possess limited computational prowess. Compounding these issues is their unpredictable availability, with a propensity to experience failures\cite{ghemawat2003google}. Thus, it becomes paramount for us to consistently oversee their performance, ensuring they offer a dependable service despite their limitations.

\noindent\textbf{Decentralized Structure and Variable Performance}: The structural difference between CDNs and PCDN is palpable. CDNs adopt a hierarchical setup, enabling them to afford clients rapid 0-RTT request-response times through effective techniques such as connection reuse\cite{nikravesh2016depth}. PCDN's structure diverges significantly, encompassing a decentralized network of peers coupled with a centralized tracker. This topology, in its complexity, hosts an ever-fluctuating number of nodes. Consequently, when a user endeavors to access content from a PCDN server and the requested data isn't cached locally, the server rejects the request, preventing the efficient connection reuse.

\noindent\textbf{NAT Punch Requirement}: A majority of the Peer Servers in PCDN operate within Local Area Network (LAN)\cite{zirngibl1995larnet}, introducing unique challenges for users. For a connection to be established with these servers, users must overcome Network Address Translation (NAT) barriers. This process, however, often results in extended video playback startup delays\cite{kihei2022comparison}. Additionally, the intricacies associated with NAT punching enhance the likelihood of connection failures, further complicating the user experience.

\begin{figure}
    \centering
    \includegraphics[width=0.48\textwidth]{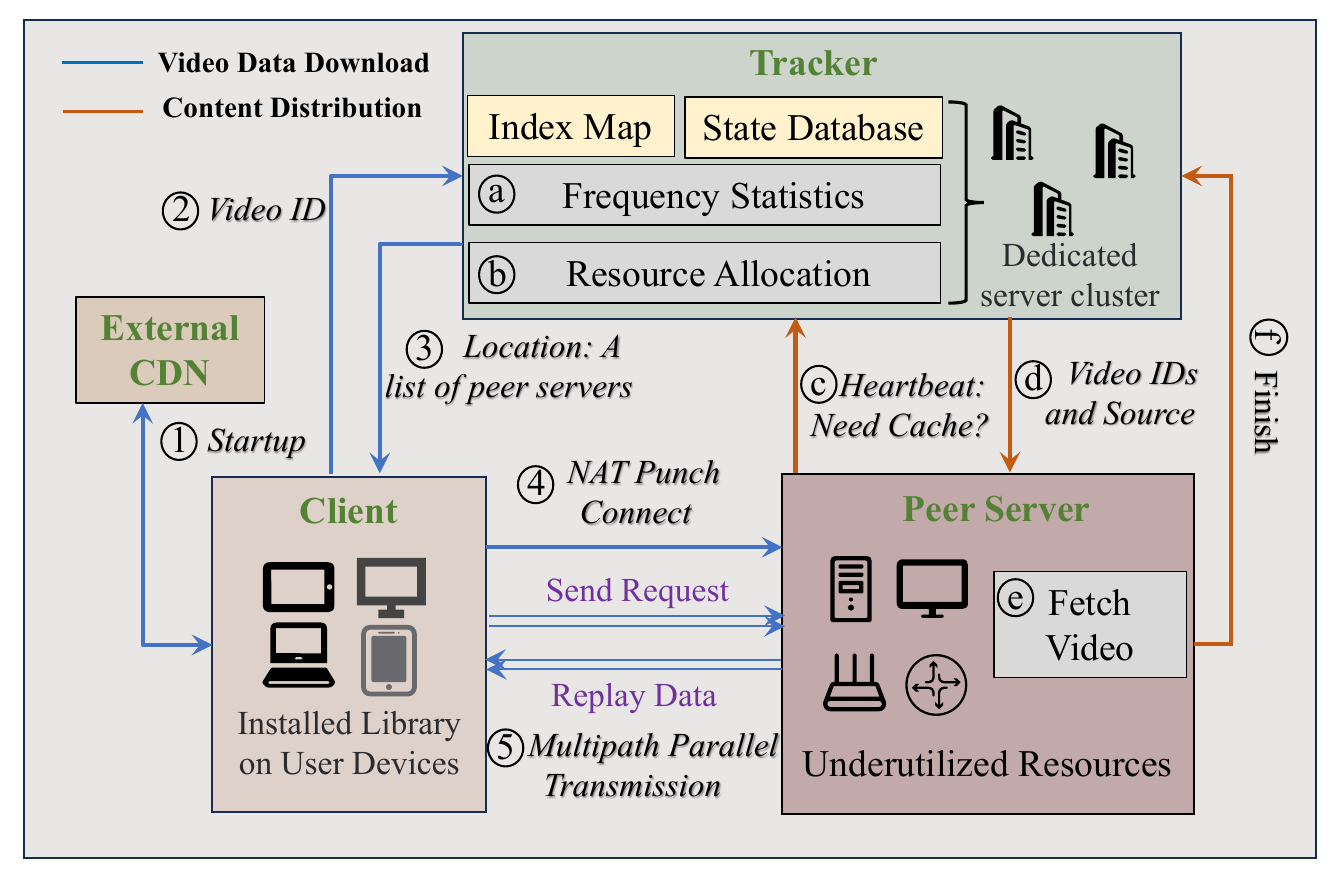}
    \caption{The \sys Overview: The four main components of \sys and its two primary flows - video downloading and content distribution. }
    \label{fig:overview}
\end{figure}
\section{Design Overview}

\autoref{fig:overview} shows the overall system architecture of \sys.
The system consists of four main components: 1) a client-side library that runs on end-user devices, 2) a tracker cluster deployed in centralized ByteDance data centers, 3) a decentralized set of peer servers with under-utilized resources running in dispersed locations, and 4) an external CDN managed by ByteDance.

Each video is uniquely identified by an \emph{ID}.
Due to the unique characteristics of video streaming, we partition each video into equal-length (in duration) \emph{segments}.
The default segment length is 10 seconds.
Video content in each segment is divided into 2 MB \emph{chunks}.
The external CDN stores \emph{all} videos.
As this paper focuses on the design of our PCDN system, we treat the external CDN as a blackbox;
readers interested in a centralized CDN design may refer to prior literature~\cite{nygren2010akamai}.

Each peer server in the PCDN caches a set of video chunks in its local memory and storage devices.
Each chunk is stored as a regular Linux file;
we rely on the OS buffer cache for in-memory caching.
These peer servers are purely passive: clients initiate video downloads from them, and trackers instruct them to store specific video chunks.
Such design minimizes the computational requirement on peer servers, a good match for their weaker and heterogeneous computational power.

% \lijl{Two questions we might want to clarity either here or in the design sections. 1) Why a peer server needs to store the entire video, not just some segments? 2) Why dividing each segment into smaller chunks?}
% \dehui{1) By doing this, Tacker only needs to store the video ID and the index between the peer server, rather than storing each segment, which greatly increases storage efficiency. Moreover, the purpose of multipath transmission is to boost aggregate throughput. If segments from the same video are spread across different peer servers, one segment would need to link to many nodes. So, when downloading the video, the client would have to connect to even more nodes compared to if each peer server had the complete video. 
% 2) A chunk is a partition on the disk, and when a video file is stored, it can only be written or read partition by partition. On the other hand, a segment is based on the idea of preloading and can be resized as needed.}

The tracker system is a centralized but distributed cluster running in our own data center.
It acts as the main control plane for \sys.
We use standard replication and partitioning techniques for fault tolerance and scaling.
The tracker system has two main sub-components.
The first component servers as a ``frontend'' metadata store for end users.
It maintains a table that maps video ID to a list of peer servers that cache the video.
User devices query this tracker component when downloading video segments from the PCDN.
The component also tracks user query statistics for analyzing content popularity distribution.
The second tracker component manages the PCDN peer servers.
It operates a database that maintains peer server states, such as their geographic location, bandwidth consumption, and disk utilization.
The component leverages detailed peer information for making fine-grained resource allocation and video distribution decisions.

The client-side library is linked into each video streaming application.
It manages video transmission scheduling, peer server filtering, pre-loading, and transport-level flow and congestion control.
The library is responsible for ensuring the video playback quality delivered to users.

\paragraph{Video downloading.}
\sys uses three main techniques to deliver high-quality video streaming experience.
First, to minimize initial video loading time (a critical criterion for user experience), the client library downloads the first few video segments from the external CDN.
The approach eschews the slow connection setup time in a P2P network, while still benefiting from the cost-efficiency of PCDN for the bulk of video transmission.
Second, to mask the low, unreliable peer server performance and bandwidth, \sys deviates from traditional single-source transmission schemes.
The client library establishes connections to multiple peer servers storing the target video.
It then schedules concurrent data transmission to achieve download speed equivalent to the aggregate bandwidth of all connected peers, akin to multi-path TCP.
Lastly, \sys applies a pipelined transmission scheme with a prefetching depth of one.
While playing the current video segment, the client library preloads the next (and only the next) segment.
The scheme minimizes both playback pauses, as the preloaded segment is likely ready when the previous one finishes playing, and wasted transmission, as clients may abandon the current video at any time.

%For video downloading, PCDN employs a multipath parallel transmission protocol an approach tailored to compensate for the limitations of individual peer servers within the network. Given that these servers typically exhibit subpar performance, are prone to failures, and individually might not meet client playback requirements, the multipath strategy becomes essential. The servers involved in a single data transfer operate independently and are unaware of each other. This necessitates localized control of transmission at the client side, managed by a specialized SDK.The video is divided into equally-sized segments for download, preloading one segment at a time. Compared to preloading the entire video, this reduces preload waste. Additionally, it facilitates decisions on which segment to switch between CDN and \sys. 

We now illustrate the video downloading process, as shown in \autoref{fig:overview}.
\Circled{1} The client library issues requests to the external CDN to download the first few segments of the target video (in a pipelined fashion).
The library stops downloading from the CDN once it establishes connections to the PCDN peer servers.
Details of the switching process are described in \autoref{CP}.
\Circled{2} In parallel to the first step, the library issues a request to the PCDN tracker cluster.
The request contains the ID of the target video.
\Circled{3} The tracker queries the content distribution map, and returns a list of peer servers storing the video segments to the client.
\Circled{4} The client establishes connections to the peer servers in the returned list.
As peer servers are in their private network, connection establishment requires NAT punching.
\Circled{5} The client library starts downloading the remaining video segments in the same pipelined style.
It receives video chunks concurrently from all connected peers using a multi-path transmission scheme.
We discuss details of this transmission scheme in \autoref{multipath}.

\paragraph{Content distribution.}
Addressing the decentralization challenge in PCDN necessitates accurate, timely resource allocation and content distribution.
\sys leverages the global vantage point of the centralized tracker system to make optimized distribution decisions.
The tracker collects real-time statistics about video access popularity, as well as peer server capacity and status.
Through an online optimization process, the tracker readjusts the target video caching distribution, allocating more aggregate bandwidth to popular videos.
Trackers, however, do not actively initiate video content redistribution.
Instead, they piggyback video caching commands when peer servers report status periodically.

%\sys employs two main techniques to dynamically adjust file deployment based on real-time popularity, ensuring that \sys has sufficient quota to handle large-scale requests.
%The ``frontend'' tracker periodically adjusts the distribution and location of video files within the network.
%The default period is 5 minutes.
%The tracker collects the number of times each video file is accessed in each interval as a measure of the video's popularity.
%Based on this popularity, it calculates how many copies of the video should be stored in each region.
%Second, the tracker implements a “passive” distribution mechanism, meaning it waits for peer servers to actively inquire about distribution results.
%The tracker informs the peer server of the videos that need to be stored and instructs it to fetch the video either from an external CDN or other peer servers, basing this decision on the peak bandwidth of the two options at that time to control costs.

This content distribution process is also shown in \autoref{fig:overview}.
\Circled{a} The tracker records per-video download frequency over five-minute intervals, using it as feedback metric to gauge video popularity.
\Circled{b} The tracker tallies the available peer server resources in each region.
This information is combined with video popularity to derive the optimal video caching distribution, both regionally and per-server.
%Then, utilizing data that captures the real-time popularity of videos, the geographical locations, and the capabilities of peer servers, the Tracker decides on the number of unique videos each region should store and designates the specific peer server for storage.
\Circled{c} Periodically, each peer server sends a heartbeat to the tracker, reporting its status.
\Circled{d}  When replying to a heartbeat, the tracker piggybacks a list of new video IDs that the peer server should cache.
For each video, it also includes the preferred downloading source --- either the external CDN or another peer server.
%check with the tracker to see if they need to fetch any new videos.
%Given tracker is the WAN cluster, it can be easily accessed by peer servers.
\Circled{e} The peer server fetches the video from the source, and stores the video locally.
\Circled{f} After the download completes, the peer server sends a confirmation back to the tracker.
The tracker then updates the video ID to peer server map.
We discuss details of this content distribution process in \autoref{scheduling}.

\section{Design Details}

\subsection{Client Library} \label{sdk}
The client library, embedded within client devices, plays a pivotal role in orchestrating data transmission control, leveraging a unique multipath parallel downloading mechanism.
The core design philosophy behind \sys's transmission is twofold: to guarantee successful data retrieval and to ensure seamless video playback.
Given the non-additive nature and simplicity between peer servers, the client library initiates all transmission and reliability operations, include: \Circled{1} Segment selection between CDN and PCDN: To ensure rapid playback initiation and fault tolerance, \sys must deftly toggle between external CDN and PCDN (\autoref{CP}).
\Circled{2} Building connections with specific peer servers: The client library queries the Tracker for indices of relevant peer servers and then initiates connections.
Given that most servers within the PCDN are on a Local Area Network (LAN), NAT punch is needed during this process.
\Circled{3} Congestion control and packet scheduling: Ensuring rapid and orderly data arrival during parallel downloads is paramount for a smooth user viewing experience.

% Key details are elaborated upon in the sections below.

% The transmission control mechanism of PCDN is implemented in the user-space based on UDP. There are two primary reasons for this decision:
% 1) The extensive transmission controls of PCDN, including features like multi-path parallel transmission, are easier to deploy and update in the user-space.
% 2) Implementing NAT is more straightforward with UDP.

\subsubsection{Hybrid CDN-PCDN Transimission}\label{CP}
The decision to switch between the external CDN and PCDN arises in two primary scenarios.
The first scenario occurs during the startup phase, from the external CDN to PCDN.
PCDN's connection process is more complex, requiring additional steps such as NAT punch and data location retrieval from the tracker.
Consequently, this complexity might lead the client to succeed in establishing a connection only after multiple attempts.
This complexity causes the PCDN's connection time to be roughly 30\% longer than a standard CDN.
To mitigate this delay and ensure rapid video startup, videos commence downloading via an external CDN.
In parallel, the PCDN initiates its connection processes.
Once the PCDN connections are ready, the client library undergoes the following checks to determine if it should transition into PCDN: \textit{PCDN reports errors?$\stackrel{N}{\longrightarrow}$ Too litte data to download?$\stackrel{N}{\longrightarrow}$ User bandwidth is too low?$\stackrel{N}{\longrightarrow}$ PCDN connection timeout?$\stackrel{N}{\longrightarrow}$ PCDN ready and buffer enough?$\stackrel{Y}{\longrightarrow}$ PCDN downloading.} 

The second scenario is the external CDN functions as a backup. It's important to know that PCDN's peer servers and network can be more susceptible to disruptions than the dedicated servers. In case PCDN encounters a failure, the client will switch to external CDN for downloading. The client ascertains potential PCDN transmission failures based on conditions through: \textit{The data volume in the buffer is below the threshold? Download rate is too low? PCDN reports errors?} If any of these conditions are triggered, the download process swiftly reverts to the CDN.

The client library issues download tasks at the granularity of segments. When a switch between CDN and PCDN is needed, the SDK will mark a specific segment to indicate its download network. This provides a buffer between the decision-making for the switch and the actual network transition, allowing for a smoother user experience.

\subsubsection{Establishment and Termination of Connections} \label{Establishment}
After the client sends a query to the Tracker, the Tracker responds by providing a list of servers that store the requested video data. Subsequently, the client initiates connections to the identified servers. During this connection phase, the client assesses various parameters from different peers, such as Round-Trip Time (RTT) and packet loss rate. Peers that exhibit unsatisfactory performance metrics, such as an extended RTT or a high packet loss rate, are discarded. Simultaneously, to communicate with peers on LAN, NAT is executed during the connection establishment phase. 
Upon completion of the video download, the connection isn't immediately severed. Instead, it's retained for a short duration. If the subsequent video is still located on the current peer, the pre-existing connection can be reused, enhancing efficiency. However, this retention period is brief, capped at around 2 minutes, given the relatively low hit rate within the PCDN.

% \textbf{Network Address Translation (NAT) Punch:}
% In the context of PCDN, edgeservers are often behind NATs, making direct client connections challenging. NAT essentially translates a device's private network address into a public one. There are four main types of NAT: Full Cone, where both IP and port are unrestricted; Restricted Cone, with only the IP being restricted; Port Restricted Cone, where both IP and port are limited; and Symmetric, which maps each session to a unique port. To overcome these NAT traversal issues, techniques like Hole Punching are employed. The essence of hole punching is for two devices behind NATs to establish communication with a known third-party server. This server aids both devices in learning about their respective public-facing addresses and ports. Once this information is exchanged, both devices attempt to establish a direct connection. The process exploits the temporary "holes" or allowed routes created in the NAT by the initial connection to the server. These methods help devices discover their public-facing addresses and establish a direct connection by exploiting the temporary routes created in the NAT, ensuring uninterrupted data downloads.
\begin{figure}
    \centering
    \includegraphics[width=0.4\textwidth]{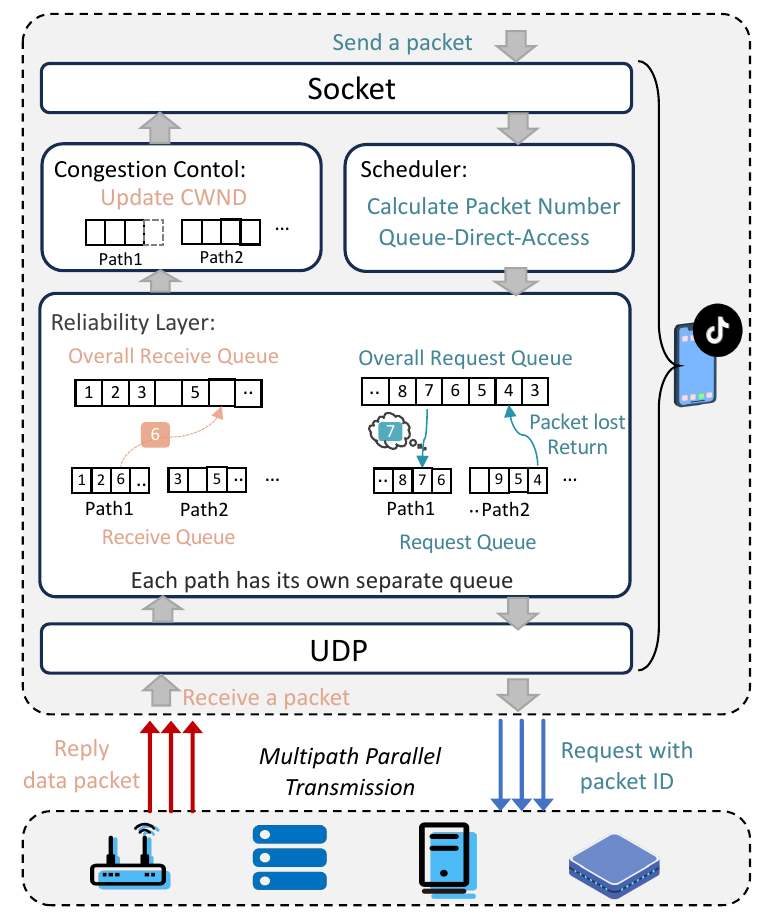}
    \caption{The Design of PCDN Multipath Transmission.}
    \label{fig:multipatht}
\end{figure}
\subsubsection{Multipath Parallel Transmission} \label{multipath}
To offer a user experience akin to that of a CDN, PCDN employs multiple edge devices to cater to a single user via parallel transmission. This results in a \emph{Multiple-Server-to-One-Client (MS2OC) pattern}, marking a notable departure from traditional multipath transmission approaches like MPTCP and MPQUIC, which adopt a One-Server-to-One-Client (OS2OC) pattern. Within the MS2OC pattern of ePCDN, the peer servers involved in a single transmission function independently, with all transmission control centralized in the client library.  In response to this design, PCDN introduces a pull-based multipath parallel transmission protocol wherein the client sends a \texttt{request} to peer servers, and in return, the servers reply with the requested data, as depicted in \autoref{fig:multipatht}. This protocol is implemented in the user-space atop UDP. Being a user-space protocol allows for easier and quicker updates and integrations as an application component, facilitating rapid changes to algorithms. Furthermore, the choice of a UDP-based transmission protocol enhances the success rate of NAT traversal. Consequently, PCDN must ensure data reliability, encompassing features like packet retransmission, congestion control, and multipath scheduling.

%This system must be capable of handling concurrent requests, efficiently distributing data loads across different paths, and ensuring the integrity of the data being transferred: 

\textbf{Data Reliability Transmission}: With the absence of any intrinsic data delivery success assurance by peer servers, the onus is on the client library to monitor data losses from the same or a different peer servers and subsequently request missing data. This reliability is facilitated by the \texttt{Request} queue and the \texttt{Received Data} queue. Each sequence number in the request queue corresponds to a single data packet within a video. The client library maintains two overall queues: \textit{1) Overall Request Queue}: Stores the sequence numbers of all yet-to-be-sent request for a video. \textit{2) Overall Received Queue}: Stores the sequence numbers of all packets received from different paths. Each path also maintains two analogous queues: \textit{1) Path Request Queue}: Captures the sequence numbers of all requests that have been sent via this path but have not yet been received. \textit{2) Path Received Queue}: Contains all packets that have been received via this path. Each path extracts the requests it needs to send from the overall request queue, sends them, and subsequently removes those requests from the overall queue. Meanwhile, the specific path's request queue will store this sequence and initiate a timer. If the corresponding packet isn't received within a stipulated time frame, the packet is deemed lost. It is then removed from the path request queue and reinserted into the overall request queue for future redistribution. 

This mechanism ensures that lost packets are retransmitted via more efficient paths. Upon the reception of a packet, its request number is deleted from the path request queue and the packet added to both received data queues. The path received data queue is designed to log packet losses, RTT, and other pertinent data, serving as an input for the congestion control algorithm. Meanwhile, the Overall Received Data Queue forwards the in-order received packets to the player for playback. Collectively, these systems ensure the dependable reception of data.

\textbf{Congestion Control}: 
In PCDN, while no specific congestion control (CC) algorithm was designed, it's recommended to use TACK\cite{li2020tack} in conjunction to address the "internal interference" challenges. Notably, most users and peer servers connect through Wireless Local Area Networks (WLAN), with WiFi as the main medium. Given WiFi's half-duplex nature and its collision avoidance mechanism, there's a direct resource contention between data requests. In practice, PCDN aggregates multiple data packet requests into a single bundled request, essentially asking for a group of packets in one go. This approach significantly improves the bandwidth efficiency during wireless transmissions. By doing so, the bulk of the transmission resources are allocated to actual data, boosting the overall efficiency of the system.

\begin{figure}[!t]
	\centering
	\includegraphics[width=0.48\textwidth]{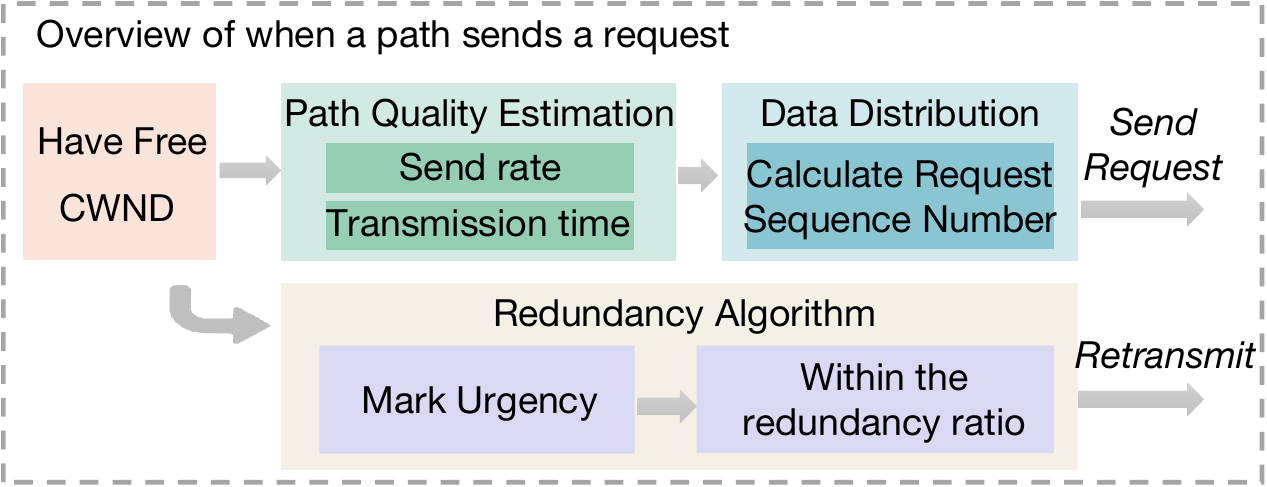}
	\caption{The Design of Multipath ByteScheduler.}
     \label{fig:scheduler}
\end{figure}

\textbf{Multipath Scheduling}: 
\sys architecture presents unique challenges in multipath scheduling for orderly data delivery.Videos that receive out-of-order packets cannot be properly played back, resulting in rebuffering and compromised playback quality. Traditional MPTCP schedulers are ill-suited for \sys. With a smaller buffer size, these schedulers halt transmissions over slower paths to maintain data sequence, leading to reduced aggregate bandwidth. Conversely, with a larger buffer size, they incorrectly assume the absence of head-of-line blocking, sending continuous data packets. This approach exacerbates data disorder, thereby increasing end-to-end latency. Thus, MPTCP schedulers find themselves in a predicament: they can prioritize either high aggregate bandwidth or low end-to-end latency but struggle to optimize both simultaneously.

MS2OC establishes a \emph{pull-based transmission pattern} where the client orchestrates singular video data \texttt{request}s directed to a group of peer servers. Each data response therefore only opens a \emph{single} flow control window in \emph{one} peer server connection. This flow control paradigm stands in contrast to MPTCP (and OS2OC broadly), where a single acknowledgment can release multiple windows due to aggregated acknowledgments. This stark difference severely limits the flow selection options available to the client. Without pausing the transmission after a data response, the scheduler will converge to a round-robin policy.

To tackle these complexities, we design a specialized packet-level scheduler for \sys, described as \autoref{fig:scheduler}. It first accurately determines the sequence number of packets suitable for the current path by a CC-decoupled prediction. It then employs the "Queue-Direct-Access" method to fetch packets from the comprehensive request queue for transmission. To enhance download speeds, the redundancy algorithm assesses if there's available space at the conclusion of each range.

\subsection{Tracker} \label{tracker}
In a PCDN architecture, clients and edge servers are in an invisible state to each other, necessitating a bridge to locate the client and the server storing specific video files. The Tracker is a substantial server cluster, with distinct subsets of servers responsible for two primary modules: video indexing and peer server management. It exposes a virtual IP address and distributes incoming requests to specific servers via a gateway. For peer management
The primary functions of the tracker include managing the index between videos and servers, planning the resources provided by vendors, and scheduling video files. We will now discuss each of these topics in detail. 

\subsubsection{Bandwidth Resource Management} \label{BRM}
For video distribution, the tracker undertakes the task of partitioning bandwidth resources from across the country. It provides a bandwidth resource guideline for different priority businesses in each region, ensuring prevention of resource contention both inter-business and inter-regionally. This also serves as a directive for subsequent file distribution.

In PCDN, bandwidth resources are provided by edge servers belonging to different vendors. The role of resource management is to allocate the limited resources among different businesses such as Douyin, Xigua, Xianshiguang, among others, while prioritizing the businesses with higher priority. To better align the distributed bandwidth resources with the businesses, these businesses are divided into different domains. Each of these domains has its own set of priority levels, and businesses with similar priorities are grouped together within them. 
Tracker evenly allocates bandwidth resources to business domains based on priority through specific steps:
\begin{itemize}
    \item \textbf{Decomposing Business Requirements:} The evaluation of bandwidth requirement $bw\_need_{rs}$ for business $s$ in region $r$ is calculated as the  the fluctuation coefficient of $s$ multiplied by its historical peak bandwidth in $r$. The capacity that can be allocated to $s$ in $r$ is $totalbw\_provide_{rs} = bw\_need_{rs} / bw\_need_r * cap_r$. Here, $bw\_need_r$ represents the total estimated bandwidth requirement for all businesses in $r$, and $cap_r$ is the calibrated bandwidth available in that region.
    \item \textbf{Decomposing Vendor Allocation:} In a given region \( r \), the bandwidth \( bw\_provide_{rvs} \) allocated by a vendor \( v \) to a specific business \( s \) is determined by \( bw\_provide_{rvs} = cap_{rp}/ cap_r * totalbw\_provide_{rs} \). Consequently, the overall bandwidth \( bw\_provide_{rvd} \) that the \( v \) can allocate to a business domain \( d \) within \( r \) is the sum of \( bw\_provide_{rvs} \) for all the businesses in this domin. Therefore, the expected bandwidth $Expbw_{rvd}$ that $v$ can allocate to $d$ in $r$ is given by $Expbw_{rvd} = Expbw_d * bw\_provide_{rvs}/ bw\_provide_v$, where $Expbw_d$ is derived based on last month's business data and the company's future growth expectations for that business.
    \item \textbf{Bandiwidth Resource Allocation: }Using a greedy algorithm, we match bandwidth resource instances to corresponding business domains. The algorithm prioritizes resource allocation to business domains with higher priority.
\end{itemize}

\subsubsection{Video File Distribution} \label{scheduling}
Every time a client queries an index tracker server for the location of a video ID, the index tracker server increments the view count for that video by one. This count is tallied periodically and reset at the start of the next cycle. Videos that surpass a specific view count threshold within the cycle and demonstrate an increased count compared to the previous cycle are allotted additional storage copies. The index tracker servers convey this data to the decision tracker servers to trigger video file distribution strategies. Video file distribution is prioritized within the same region because the popularity of a video might differ across regions. Distributing predominantly within a particular region minimizes inter-region transfers, thereby reducing latency. Taking a region as an example, the tracker first checks the bandwidth resources allocated to that region. It then calculates the proportional popularity of the video in relation to the overall popularity within that region. This proportion is used to determine the bandwidth resources that can be earmarked for distributing this particular video, subsequently ascertaining the number of extra copies the video can be granted.

Once the tracker finalizes its decisions, it awaits peer servers to retrieve the results. At the end of each cycle, every peer server polls the tracker to inquire if there are files designated for storage on its end. The tracker then informs the peer server about the video IDs it needs to store and the source location (an external CDN or other peer servers) from which these videos should be downloaded. 
If both CDN and other peer servers in PCDN have the requirement video, peer servers have the option of retrieving this data either from the CDN or another peer, with speed prioritized during peak hours and cost-effectiveness during off-peak times. 
Upon successfully obtaining the video from the specified location, peer servers send a confirmation back to the tracker, marking the successful completion of the distribution process. This sequence constitutes a single file distribution cycle.

In PCDN, the file distribution cycle is set at 5 minutes and the viewing threshold is 100. According to practical experience, this real-time strategy of adjusting popularity scores and distributing files is more effective at adapting to the rapidly changing nature of video popularity than deploying files at fixed times daily. It allows for prompt increases in file coverage and boosts the PCDN sharing ratio in cases where a video suddenly gains popularity. However, this operation is halted during peak hours, unless a video becomes exceedingly popular, to control peak bandwidth usage and reduce costs.

\subsubsection{Index Management and Peer Allocation}
Every time a peer server successfully stores a video file, it notifies the tracker. As a result, the tracker maintains a comprehensive index of $<VideoID, Peer\_Server\_List>$, ensuring it has a real-time understanding of which videos are stored on which servers.
When a client wants to download data from servers, it first requests tracker for available download points. Tracker initiates the process by querying a sufficient number of $n$ nodes. Subsequently, the tracker scores these nodes and then returns the $m$ nodes required by the client. The scoring is based on various criteria including bandwidth utilization, distance, CPU utilization, and NAT matching.

\subsubsection{Peer Server State Management}
Upon startup, each peer server reports its current region, IP address, and associated vendor to the Tracker. Subsequent changes to these details are immediately upload. Moreover, peer servers send performance metrics like disk and bandwidth utilization to the tracker at 30-second intervals. This real-time data helps the Tracker in video data scheduling and distribution. If a peer server fails to report within the scheduled period, the Tracker assumes the server may be offline and removes it from the active peer pool as a maintenance measure. 

\subsection{Peer Servers} \label{peer}
PCDN utilizes a chunk-based storage methodology for video data on every peer server. Each chunk is housed as a standard Linux file, leveraging the OS buffer cache for in-memory storage. While videos are segmented into chunks, there's a bias toward keeping an entire video file on a single peer server. This approach streamlines management and reduces the number of peer servers needed for a single transfer. Each chunk is paired with a unique checksum value, bolstering its integrity. This system not only confirms the accuracy of each segment but also provides an added layer of protection against potential discrepancies that may arise during inter-peer server transfers.

\subsubsection{Video Storage Replacement }
Whether downloading or uploading, video data must be stored locally on servers. As each server has a limited storage capacity, when a new popular chunk is allocated to a server without sufficient storage, some chunks must be replaced. Common server storage replacement strategies primarily include replacement based on video access time and replacement based on video access frequency, among others. However, neither of these methods takes into account the video status and the chunk's position. Moreover, in practice, the popularity of videos frequently changes, requiring constant adjustments in video distribution. Thus, the strategy for video chunk replacement is crucial, aiming to reduce the number of video replacements and increase the hit rate. Therefore, PCDN evaluates each chunk by considering both the video status and the position of the corresponding chunk. The chunk with the lowest score is then chosen for replacement.

To address these challenges, PCDN implements a nuanced replacement strategy that prioritizes videos for replacement based on several criteria: 1) Videos not currently being transmitted have a higher replacement probability than those in transit.
2) Less frequently accessed videos are more likely to be replaced than popular ones.
3) Videos only stored in disk have a higher replacement probability than those cached. The highest priority videos get replaced first. In cases where multiple videos share the same priority, the system prefers replacing videos that are of similar size to the incoming video. If the selected video for replacement is smaller than the incoming one, the system will look for another candidate. 
When new videos are introduced to the peer sever, the aforementioned procedure is iteratively applied to each video until they are all successfully integrated. The addition and removal of videos are performed chunk-by-chunk within the video file. Therefore, prior to initiating the replacement, a temporary file is generated, documenting the modified chunks and their respective sequence numbers. If any issues arise during the operation, the content can be restored using the data in this temporary file.

\subsubsection{Unified Peer Servers Management Platform}
PCDN has implemented a PaaS platform to centrally oversee and manage all peer servers, ensuring consistent business stability across heterogeneous devices. Utilizing peer servers from various manufacturers and device types to deliver video services introduces several challenges: 1) Diverse Device Resources: The vast differences in device resources mean businesses need to actively adapt to various device environments. This process prolongs development cycles, making tasks more cumbersome and extensive. 2) Inconsistent Vendor Technologies: Entrusting deployment and updates to manufacturers with varying technological expertise can be risky, leading to potential system instability. 3) Lack of Resource Coordination: Manufacturers may arbitrarily install firmware packages, dictating the operational scale. This makes it difficult for the company to control costs. 

To address these issues, PCDN's PaaS platform offers a unified management system for all these diverse peer servers, enhancing business reliability. This platform standardizes software management for all peer servers. It can issue real-time tasks to device components, such as orchestration, updates, and file management. Furthermore, the platform can gauge the bandwidth resources provided by each vendor and report it to the tracker's resource pool for bandwidth allocation. In addition, the platform handles device maintenance, periodically conducting health checks and fault detection. It promptly issues alerts for any anomalies. As the service demands shift, the platform can also automatically scale the resources of peer servers up or down.

\section{Evaluation}
Having been operational within ByteDance for over six years, \sys has consistently maintained its performance without major declines. Yet, to elucidate the efficacy of \sys, we embarked on a large-scale A/B testing in a real-world network, juxtaposing the performance of Bytedance CDN and PCDN by running them concurrently. Engaging approximately ten million participants in each group, the tests dealt with billions of video playbacks daily.  Bytedance CDN (It is referred to as CDN later in the paper), equipped with cutting-edge technology and optimized for both long and short video services, stands among the best in the industry. However, our results indicated that PCDN could match, closely mirroring the QoE provided by the Bytedance CDN. This demonstrates PCDN's potential as a reliable content delivery solution. Unless otherwise specified, the data in the figures represent the ratio of \sys to CDN performance.

\subsection{Video Transmission Speed}
\begin{figure}
    \centering
    \includegraphics[width=0.4\textwidth]{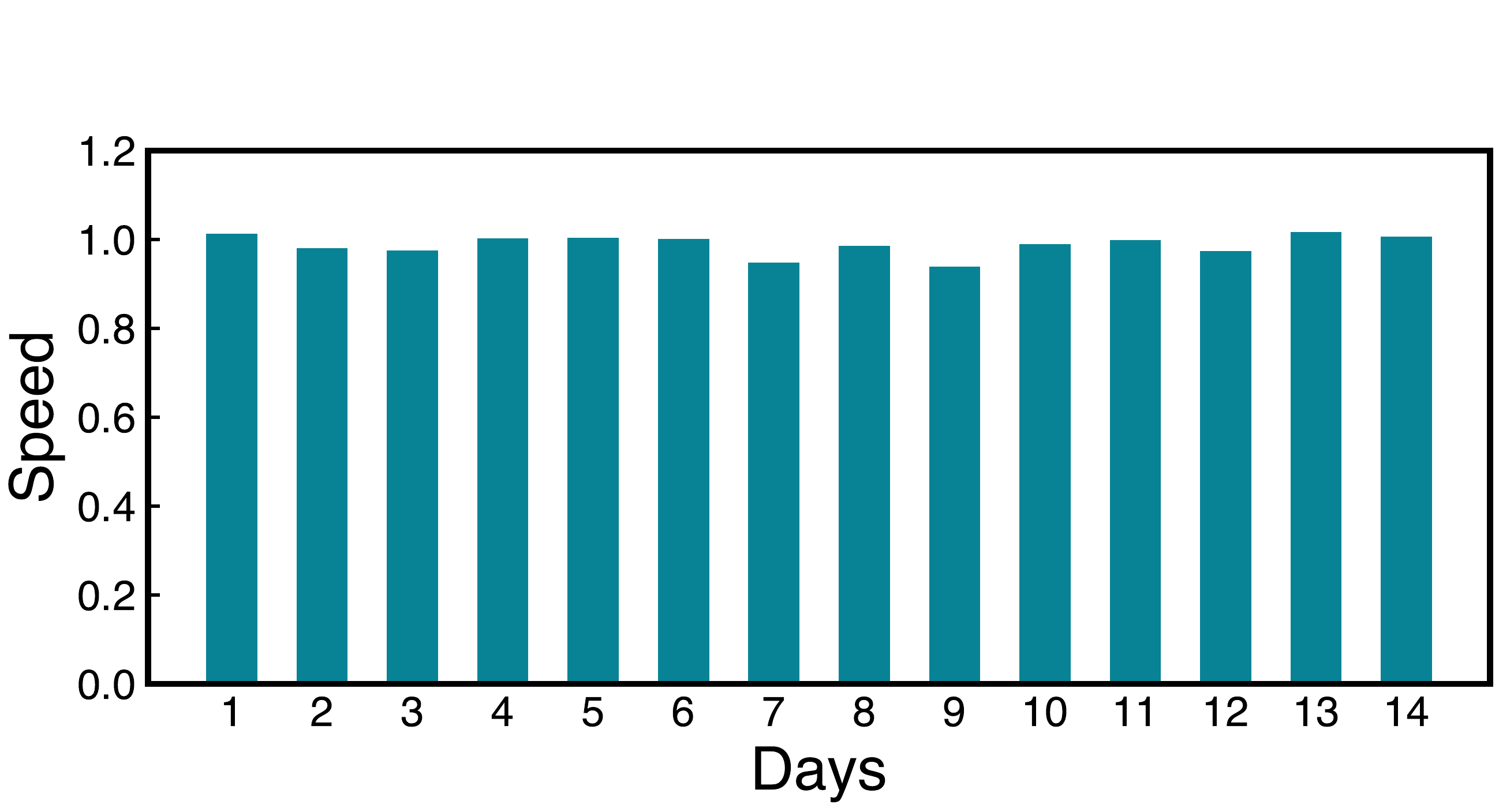}
    \caption{Results on video transmission speed. \sys achieves comparable transmission speeds to CDN, even during peak times on Saturdays and Sundays. }
    \label{fig:speed}
\end{figure}
The quality of experience (QoE) for users during video playback is intrinsically linked to the download speed. Factors such as video bitrate quality and a reduction in rebuffering rate are directly influenced by this. \autoref{fig:speed} presents the ratio of daily download speeds between \sys and CDN. To calculate this data, the daily average download speed for each video is considered. Times when no data downloads are excluded from speed calculations. For the PCDN data set, instances where there's a fallback to the CDN are filtered out, ensuring no speed enhancements from the CDN skew results. Online statistics show that the fallback rate to CDN is around 2\%, primarily due to poor client network connectivity.
The seven-day statistical data suggests that PCDN download speeds match those of CDN. Both the effective download speed and the low fallback rate benefit from multipath transmission. This transmission method compensates for potential bandwidth limitations of individual peer servers, and increasing the network's overall robustness.

\subsection{Establishment Time and Cost Saving}
\begin{figure}
    \centering
    \includegraphics[width=0.4\textwidth]{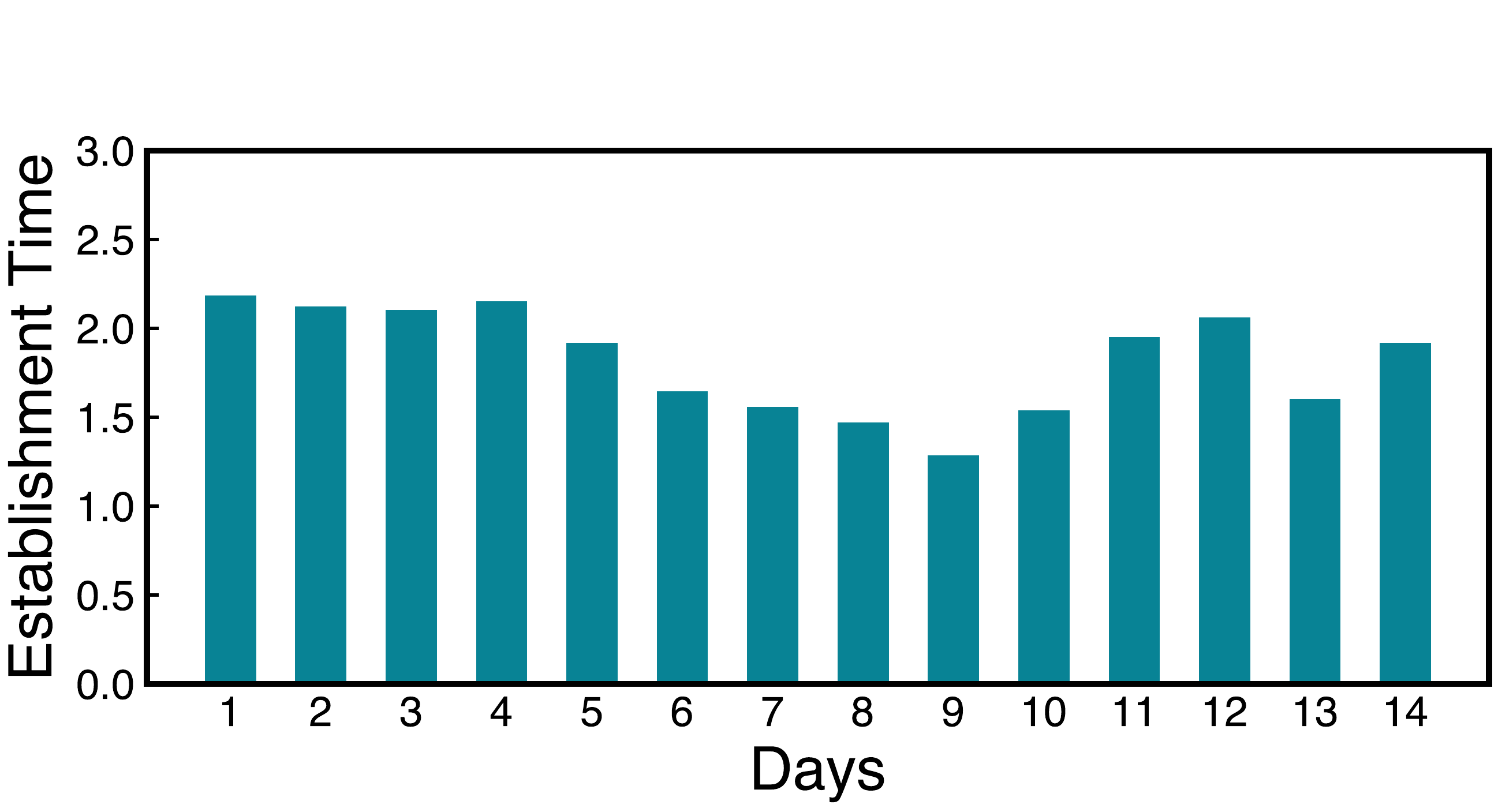}
    \caption{Results on establishment time. \sys has a slightly longer connection establishment time compared to the CDN, but does not impact the cost savings.}
    \label{fig:firstRTT}
\end{figure}
\autoref{fig:firstRTT} provides a comparative analysis of the connection establishment times between \sys and the CDN. As per our observations, \sys takes about 1.5 to 2 times longer to establish a connection compared to the CDN. At first glance, this might raise concerns. However, it's essential to contextualize this difference: given that the CDN's connection time is already extremely short, even doubling that duration only results in an additional delay of a few tens to hundreds of milliseconds. 
Empirical evidence suggests that approximately 80\% of the video data in \sys is directly retrieved from peer servers. This results in the per-unit traffic cost for \sys being significantly lower than conventional CDNs. On average, the traffic cost can be reduced by about 40\%, translating to a savings of hundreds of millions of RMB for the company. In essence, while \sys may introduce a slight initial latency, it stands out as a more economical choice. This ensures users benefit from high-quality video streaming at a reduced traffic expense.

\subsection{Video Rebuffer Rate and Time}

\begin{figure}[!t]
\centering
\subfigure[Rebuffer Rate.]{
\begin{minipage}[t]{0.21\textwidth}
\centering
\includegraphics[width=1\linewidth]{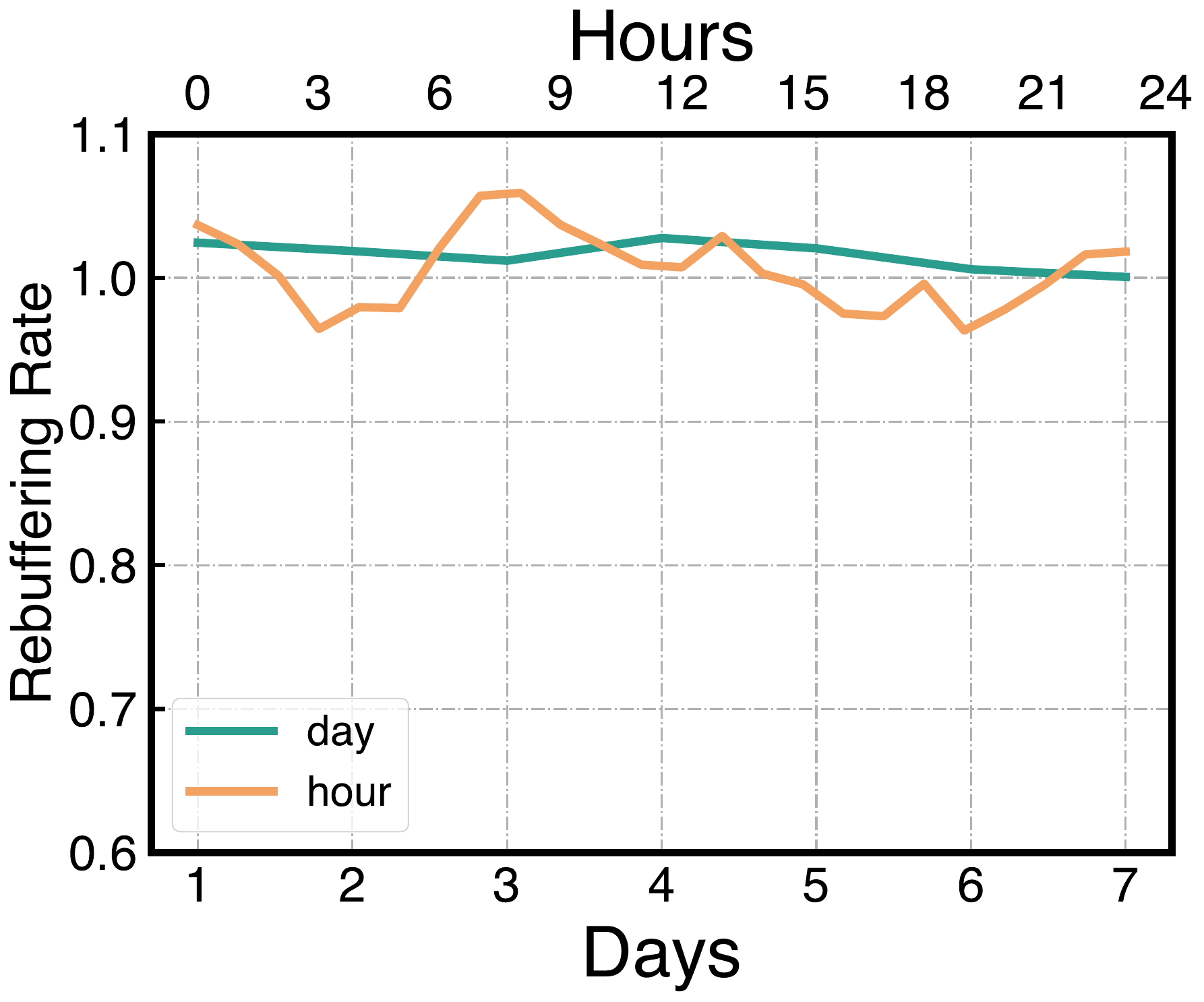}
% \caption{A/B test results on rebuffer Rate.}
\label{fig:rebufferRate}
\end{minipage}
}
\subfigure[Rebuffer Time.]{

\begin{minipage}[t]{0.22\textwidth}
\centering
\includegraphics[width=1\linewidth]{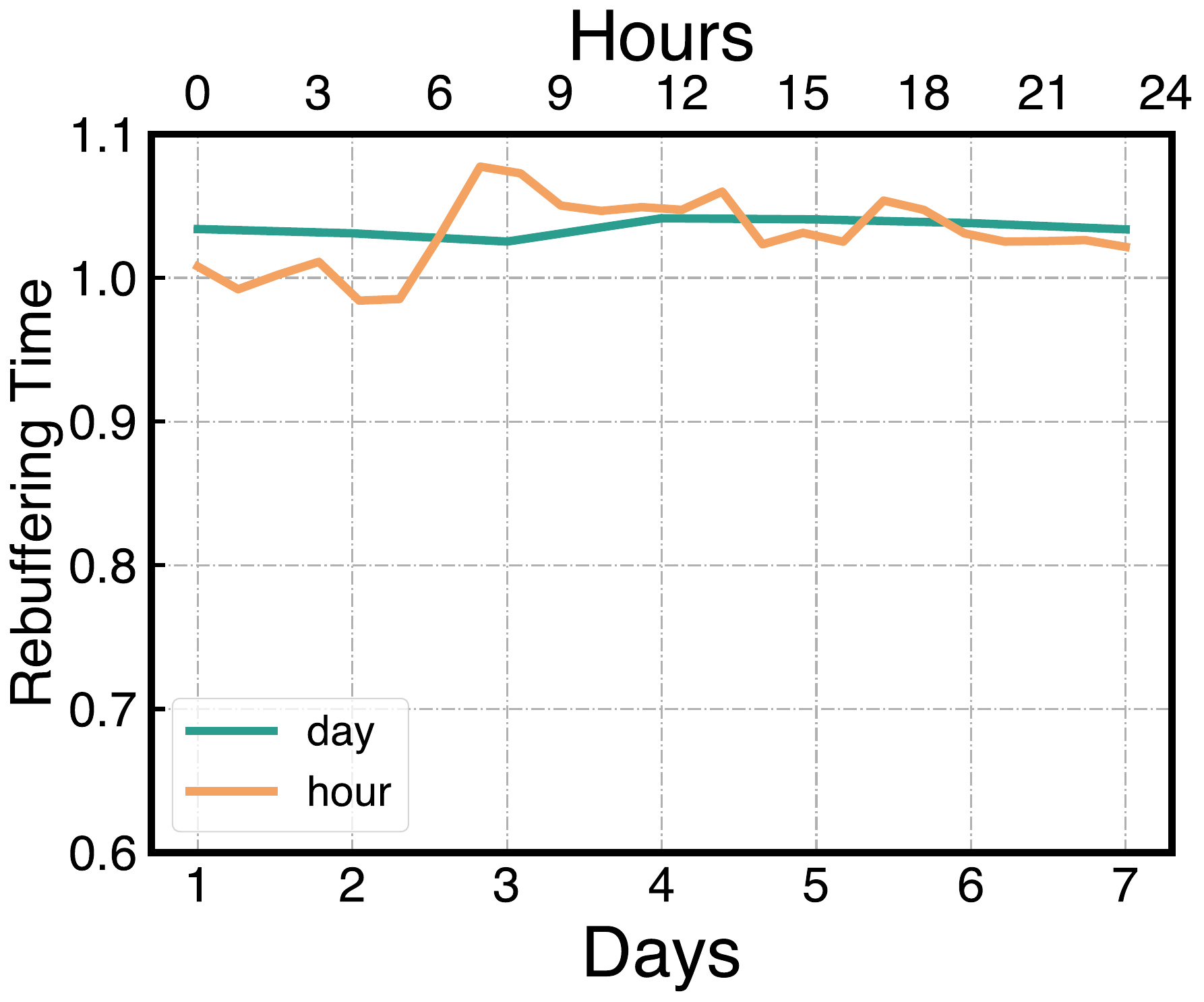}
% \caption{rebuffer Time per 100 seconds}
\label{fig:rebufferTime}
\end{minipage}
}
\caption{Results on video rebuffer. Compared to CDN, \sys shows no significant increase in rebuffer. }
\label{fig:rebuffer}
\end{figure}

\autoref{fig:rebufferRate} delineates the rebuffer rate, which is computed as the number of rebuffer events divided by the total number of videos. This metric is presented across both hourly and daily intervals. Conversely, \autoref{fig:rebufferTime} illustrates the average rebuffering duration per 100 seconds, aggregated over the same hourly and daily periods. Impressively, neither metric shows an increase when compared to CDN. This indicates that PCDN effectively harnesses these inexpensive devices to fulfill user viewing demands. Furthermore, the hourly data implies that even during peak traffic times, PCDN maintains its performance, exhibiting neither increased rebuffering rates nor durations, successfully managing the rise in peak-hour traffic.

\subsection{Segment Splitting Size}

\begin{figure}[!t]
	\centering
	\includegraphics[width=0.3\textwidth]{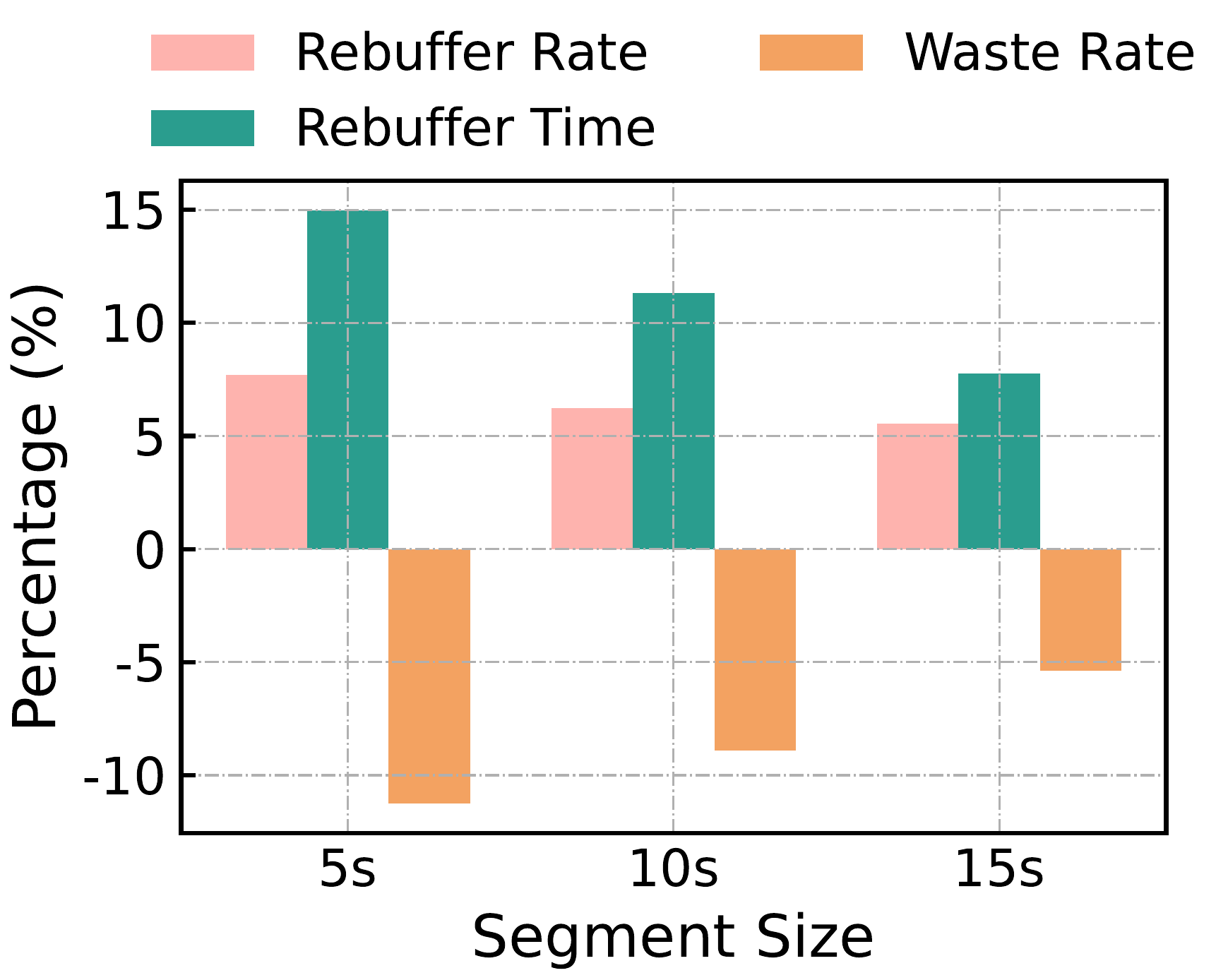}
	\caption{Testing various segment sizes to strike a balance between QoE and cost.}
        \label{fig:rangesize}
\end{figure}
\sys employs a strategy of segmenting videos into equal-sized segments for download. As users watch one segment, the system preemptively initiates the download of the next segment. This approach offers two significant advantages: 1) Minimization of preload waste. If a user swipes away or exits the video prematurely, any preloaded segments that remain unwatched result in wasted data, increasing bandwidth costs. 2) Seamless switch control between CDN and \sys: Within the duration of a current segment, \sys can dictate which network to use for the upcoming segment. 
The ideal segment size is crucial. If it's too large, there's a risk of more unused data; if it's too small, it can cause frequent video buffering. To determine the optimal size, we conducted experiments, segmenting videos into 5s, 10s, and 15s intervals, and compared these to full-video preloading. As illustrated in \autoref{fig:rangesize}, smaller segments resulted in reduced buffering and wastage. Given the need to strike a balance between cost efficiency and video playback quality, we've settled on segmenting videos into 10s chunks, providing an optimal trade-off between data costs and user experience.

\subsection{The Efficacy of Multipath Scheduler}
\begin{figure}[!t]
	\centering
	\includegraphics[width=0.4\textwidth]{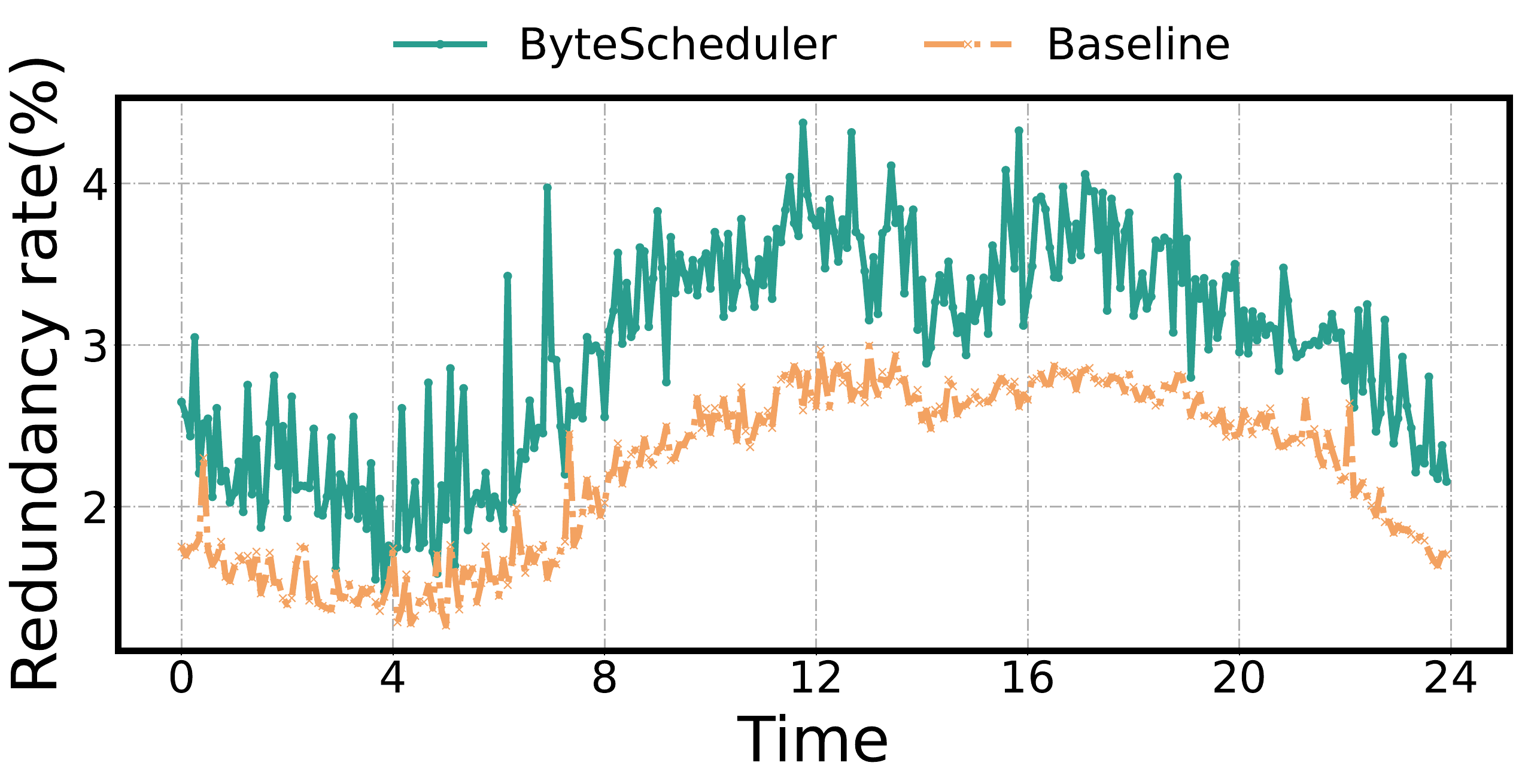}
	\caption{Results on the redundancy rate of ByteScheduler and baseline within a day.}
     \label{fig:schedule_waste}
\end{figure}

\begin{figure}[!t]
	\centering
	\includegraphics[width=0.4\textwidth]{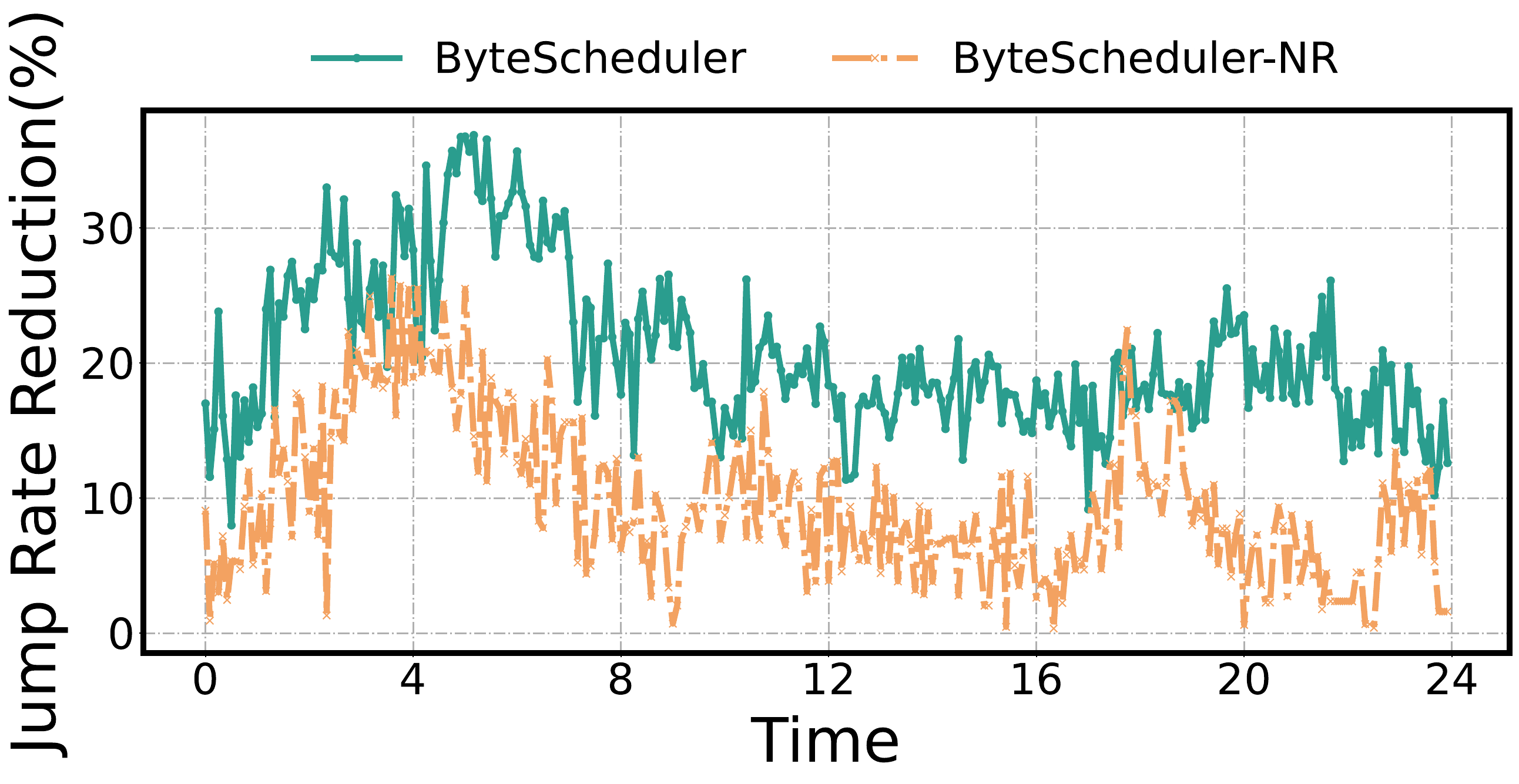}
	\caption{Results on the reduction in the rate of videos that
jump out to the CDN network due to rebuffering.}
     \label{fig:jump_rate}
\end{figure}

To measure our multipath scheduler's effectiveness, we extracted two groups from the \sys test set: one using ByteScheduler and another with a baseline algorithm that dispatches packets via the shortest RTT path. \autoref{fig:schedule_waste} and \autoref{fig:jump_rate} depict `Redundancy rate' and `Jump rate', respectively. Notably, the Jump rate illustrates the reduction achieved by ByteScheduler and its variant, ByteScheduler-NR (without tail redundancy), compared to the baseline.

\textit{Efficiency Analysis: Data Redundancy Rate Examination.}
Data redundancy rate is another pivotal metric, indicating the efficiency of data utilization. It's determined by contrasting the total packets in a task against the discrepancies between sent requests and actual packet count. As depicted in \autoref{fig:schedule_waste}, ByteScheduler's redundancy rate, while marginally surpassing that of baseline, remains minimal. This slight increase is attributed to ByteScheduler's redundancy algorithm. However, this minimal trade-off (a mere 0.74\% increase) pales in comparison to the substantial 23.93\% enhancement in video download speed. Clearly, ByteScheduler's redundancy strategy leverages only a trifling amount of surplus data but delivers paramount benefits.

\textit{Impact on Video Quality: Jump Rate Insights.}
The video jump rate is a crucial indicator of video QoE. A higher rebuffer rate can severely impact the user's viewing experience, indicating interruptions in video playback. Within Douyin's \sys system, any rebuffering incident triggers a network transition from \sys to the more reliable CDN. This measure, known as the 'online jump rate', offers insights into the instances where \sys faced transmission hiccups. \autoref{fig:jump_rate} presents a comparison between ByteScheduler and the baseline mechanism across various times of the day. Notably, the adeptness of \sch stands out as it curtails the jump rate by an impressive 30\%, enhancing the overall video streaming quality.

\subsection{Playback Failure Rate}
The playback failure rate refers to instances where users are forced to exit due to issues related to network, device, player anomalies, or system bugs. Remarkably, PCDN's failure rate is two orders of magnitude lower than that of the CDN, with a rate hovering around $10^{-4}$. While the majority of failures in both CDN and PCDN predominantly arise from user-side issues, PCDN's notably lower rate can be attributed to its inherent flexibility in scalability and its multipath transmission, offering heightened robustness against potential interruptions.

\begin{figure}
    \centering
    \begin{minipage}[t]{0.21\textwidth}
    \centering
    \includegraphics[width=1\textwidth]
        {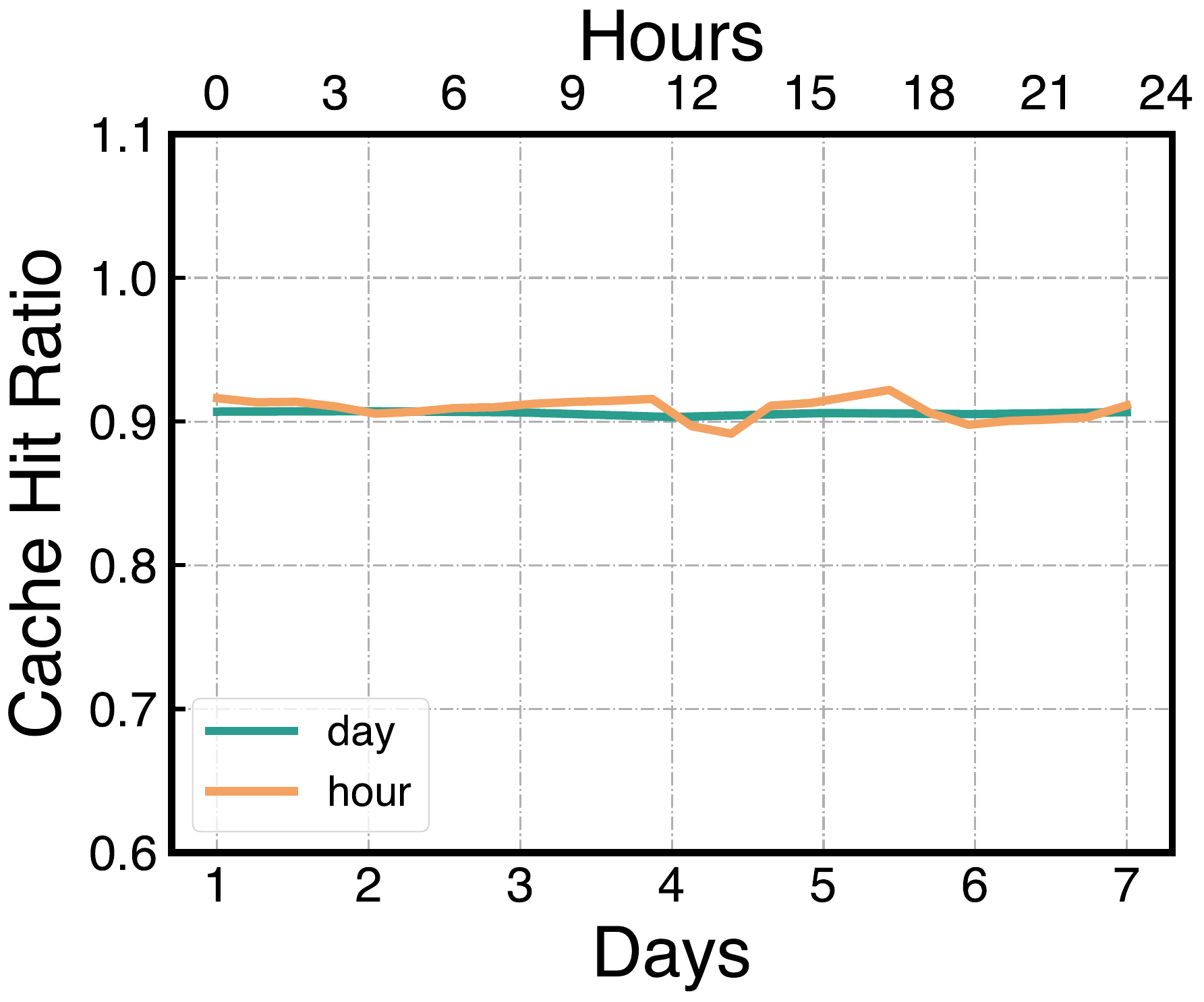}
    \caption{Results on cache hit ratio.}
    \label{fig:hit}
    \end{minipage}
    \hspace{1mm}
    \begin{minipage}[t]{0.21\textwidth}
    \centering
    \includegraphics[width=1\textwidth]       {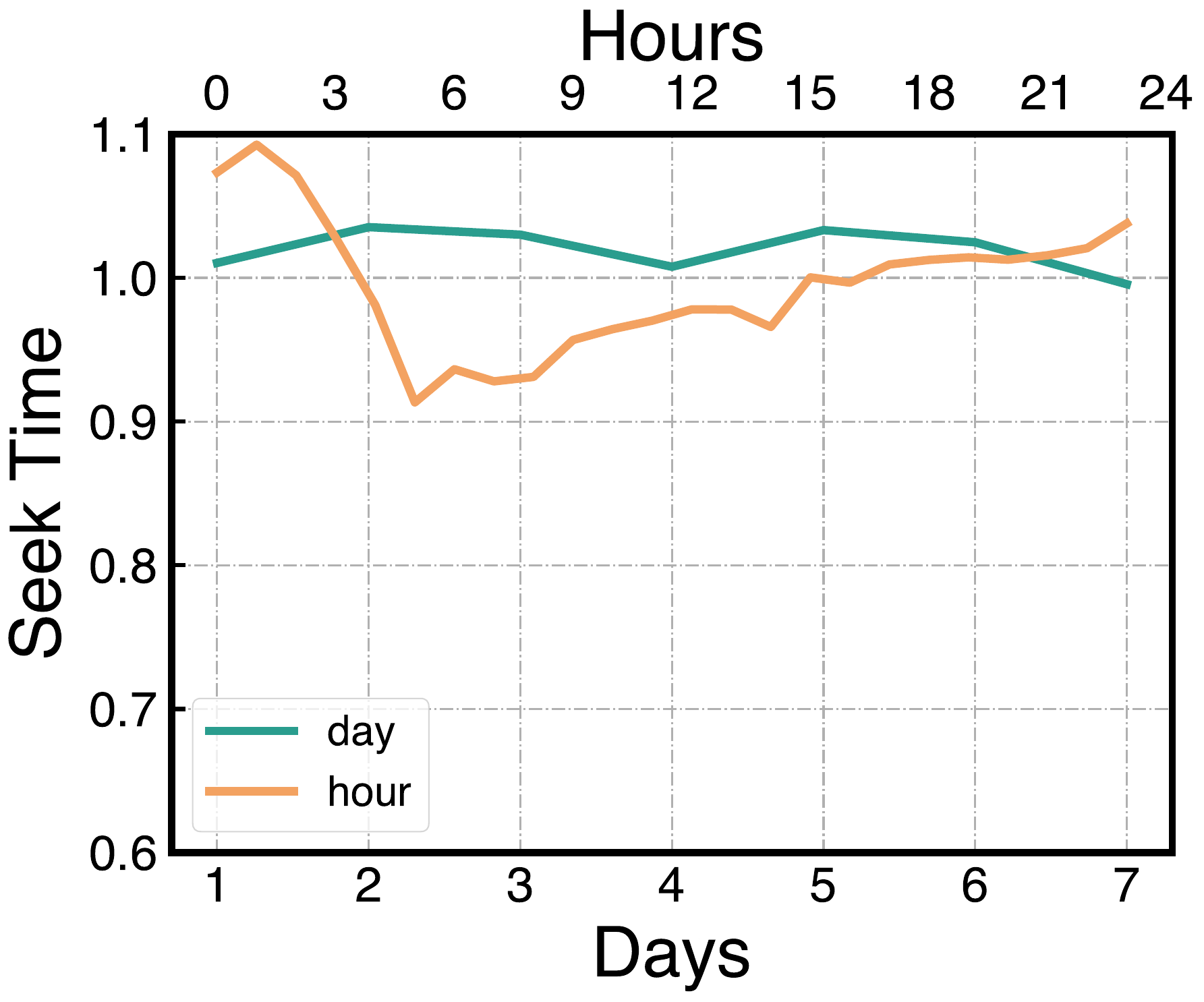}
    \caption{Results on the seek time.}
    \label{fig:seek}
    \end{minipage}
\end{figure}

\subsection{Cache Hit Ratio}

Figure 13 displays the cache hit ratio, indicating the frequency of successful file retrievals from memory or cache during downloads. PCDN surpasses CDN in this regard, thanks to its predictive deployment mechanism. PCDN constantly adjusts the “popularity to deployment volume” based on the strategy in \autoref{tracker}, ensuring each video file is optimally distributed, avoiding overloads or insufficiencies. This boosts PCDN's cache hit ratio. On the other hand, CDNs, with their centralized structure, often have to fetch videos from the primary source, resulting in a lower cache hit ratio.

\subsection{Seek Time}
Seek time refers to the average duration from dragging the video progress to the point when the corresponding image is displayed, reflecting the network's emergency response capability. When users watch videos, their behavior is exploratory and random, yet their patience for loading times is limited. Although \sys takes longer to establish a connection, it still maintains a transmission capability comparable to that of a CDN during the transfer process.

% \begin{table}[!htbp]
% \centering
% \caption{Result}
% \begin{tabular}{ |c|c|c|c| } 
%  \hline
%  Range Size & Rebuffer Rate & Waste Rate & Rebuffer Time \\ 
%  \hline
%  5s & +7.71\% & -11.24\% & +14.98\% \\ 
%  \hline
%  10s & +6.23\% & -8.91\% & +11.33\% \\ 
%  \hline
%  15s & +5.55\% & -5.39\% & +7.76\% \\ 
%  \hline
% \end{tabular}
% \end{table}

\section{Related Work}
\textbf{Distribution Network Systems:}
% CDNs have evolved as the backbone of web content delivery, strategically positioning servers across the global landscape. Introduced by IETF, the CDI (Content Delivery Inter) architecture laid foundational groundwork for CDNs\cite{bertrand2012use}, with DNS interposition playing a pivotal role in integration\cite{kangasharju2001performance,shaikh2001effectiveness}.
% Among the giants in the space, Akamai is a pioneer\cite{nygren2010akamai}, boasting a highly decentralized infrastructure that ensures expansive content distribution and robustness against disruptions. In contrast, Limelight adopts a more centralized model, deploying fewer yet powerful servers in key global cities, offering a balance between reach and power. Cloudflare\cite{website:cloud} extends its presence in over 200 cities across 100 countries, uniquely emphasizing user security and website protection alongside content delivery, positioning itself as a guardian against malicious threats. Lastly, Amazon CloudFront uniquely capitalizes on its integration with the AWS ecosystem, making it a preferred choice for businesses already embedded in the Amazon cloud infrastructure.
% Initiating a CDN demands significant investment for a global presence. Consequently, providers typically charge either fixed fees or ones based on data transfer and storage. While they provide unparalleled service quality, it often comes at a higher cost.
CDNs are the linchpin of web content delivery, with their design heavily influenced by the CDI architecture introduced by IETF\cite{bertrand2012use}. Akamai, a key player, offers decentralized content distribution\cite{nygren2010akamai}, while Limelight\cite{canali2018designing} uses fewer, stronger servers in major cities. Cloudflare focuses on security in its widespread network\cite{website:cloud}, and Amazon CloudFront leverages its AWS integration. Establishing a CDN requires vast investment, leading to either fixed or data-driven fees, ensuring top-tier service at a premium cost.

% P2P networks have risen as a decentralized alternative to traditional content delivery mechanisms. A hallmark in this domain is BitTorrent\cite{cohen2003incentives}, which optimizes bandwidth usage by dividing files into segments and leveraging a torrent file for metadata management. Earlier in the P2P evolution, Napster paved the way with its centralized server handling indexing and peer discovery, while ensuring direct peer-to-peer file exchanges. Systems like Gnutella\cite{website:Gnutella} and Freenet\cite{clarke2001freenet} took a distinct route, emphasizing decentralized approaches where search requests are broadcast over a network of peers. Further refining the model, Kazaa\cite{website:kazaa} utilized a hybrid architecture, where certain peers, termed "supernodes", took on the responsibility of content indexing within their vicinity. While P2P networks might occasionally lag behind CDNs in terms of reliability, their inherent redundancy provides resilience during high-demand scenarios. One trade-off, however, is the potentially slower propagation of new content in P2P networks, as it depends on exchanges between individual nodes.

P2P networks have emerged as a decentralized counterpart to conventional content delivery systems. BitTorrent\cite{cohen2003incentives} stands out for its efficient bandwidth use and segmented file approach. Early P2P systems like Napster used centralized servers, while Gnutella\cite{website:Gnutella} and Freenet\cite{clarke2001freenet} emphasized broad peer-based searches. Kazaa\cite{website:kazaa} introduced a hybrid model with "supernodes" handling content indexing. While P2P networks offer redundancy advantages over CDNs, they can sometimes be slower in disseminating new content due to peer-based exchanges.

\textbf{Multipath Protocol and Scheduling Algorithms:}
In recent years, as devices boast an increasing number of network interfaces, multipath transmission protocols have emerged to harness this potential \cite{ford2013tcp, coninck2020multipath, guo2017accelerating}. Protocols like MPTCP\cite{ford2013tcp} and MPQUIC\cite{de2017multipath} augment traditional single-path protocols based on TCP \cite{forouzan2002tcp} and UDP, enabling end-to-end multipath transmissions. 

MPTCP scheduling has evolved from its initial RoundRobin approach \cite{imaduddin2021multipath} to the minRTT algorithm in the Linux kernel, prioritizing paths with minimal RTTs. Algorithms like Otias\cite{yang2014out}, ECF\cite{lim2017ecf}, and BLEST\cite{ferlin2016blest} focus on optimizing fast paths, potentially reducing aggregate throughput. Redundancy strategies, such as TWC\cite{xing2021low} and XLINK\cite{zheng2021xlink}, ensure structured data transmission while addressing the HOL challenge, even if it slightly impacts overall throughput.

\sys is a video distribution system that blends a centralized control plane with a decentralized data plane. Unlike fully decentralized distribution systems, \sys has a more pronounced centralized control plane, the tracker. This tracker makes explicit decisions for all peer servers, guiding them on where to download videos and determining which videos each peer server should store. By doing so, it achieves a more optimal global view, ensuring the reliability of video services. Compared to fully centralized distribution systems, \sys leverages more dispersed residual resources, leading to cost savings. In terms of transmission, \sys employs a multi-path parallel transmission mode and has designed a unique multi-path scheduling method specifically tailored to its distinctive model.

\section{Conclusion}
\sys represents a transformation of the traditional CDN architecture, making the content distribution network more flat and scalable. It successfully organizes the diverse and arbitrary peer devices into a platform that ensures quality video service. Relying on a centralized control plane, the tracker, \sys distributes files to peer servers to dynamically adjust resources according to demand and directs clients to download locations. This design approach for peer servers minimizes their computational requirements, aligning well with their diverse and generally lower computational capacities. Meanwhile, the client handles all transmission control operations. To date, \sys has been operating successfully for six years, saving the company hundreds of millions of yuan while maintaining performance comparable to traditional CDNs.

% Primarily refers to the pre-distribution.

% How to pre-distribution: Each device will periodically query the Tracker to check if there are any files that need to be distributed to it. The download process utilizes the HTTP protocol for file retrieval.

% Download: Files allocated to devices are either fetched from a CDN or obtained from other devices. CDN downloads are fast but expensive, while P2P downloads are cost-effective but less reliable. To manage costs, different download methods are employed during various time periods. During off-peak hours, a hybrid download approach is utilized to control expenses. During peak hours, the source with the fastest download speed is selected for retrieval.

% Upload information of devices: 1. Upon successful file download, the device reports to the Tracker, indicating that the specific file is now available for use.
% 2.The device regularly sends information to the Tracker, including its operator, IP address, geographical region, and other details. Additionally, it provides updates on its bandwidth utilization, disk usage, and available bandwidth every 30 seconds.

% Delete files: When a device receives a new file for distribution but lacks storage space, it will proceed to delete the file that has been accessed the least recently to make room for the new file.

{\footnotesize \bibliographystyle{acm}
\bibliography{sample}}

\end{document}